\documentclass{aa}  

\usepackage{graphicx}
\usepackage{txfonts}
\newcommand{\kms}{\mbox{$\>{\rm km\, s^{-1}}$}}
\newcommand{\pc}{\>{\rm pc}}
\newcommand{\kpc}{\mbox{$\>{\rm kpc}$}} 
\newcommand{\kmsk}{\mbox{$\>{\rm km\, s^{-1}\, kpc^{-1}}$}}
\newcommand{\Gyr}{\mbox{$\>{\rm Gyr}$}}

\begin{document} 

   \title{Dynamical interplay of disc thickness and the interstellar gas: implication for the longevity of spiral density waves}
\titlerunning{Effect of disc thickness on longevity of spiral density waves}
   \author{
   Soumavo Ghosh\inst{1,2} \thanks{E-mail: ghosh@mpia-hd.mpg.de}
   \and 
   Chanda J. Jog\inst{3} \thanks{E-mail: cjjog@iisc.ac.in}
   }
\authorrunning{S. Ghosh and C.J. Jog}
   \institute{Max-Planck-Institut f\"{u}r Astronomie, K\"{o}nigstuhl 17, D-69117 Heidelberg, Germany
   \and
   Indian Institute of Astrophysics, Koramangala II Block, Bengaluru 560034, India
   \and
   Department of Physics, Indian Institute of Science, Bangalore 560012, India
   }
   
   \date{Received XXX; accepted YYY}

 
  \abstract
  {A typical galactic disc has a finite thickness and in addition to stars it also contains a finite amount of interstellar gas. Here, we investigate the physical impact of the finite thickness of a galactic disc on the disc stability against the non-axisymmetric perturbations and on the longevity of the spiral density waves, with and without the presence of gas. The longevity is quantified via group velocity of density wavepackets. The galactic disc is first modelled as a collisionless stellar disc with finite height and then more realistically as a gravitationally-coupled stars plus gas system (with different thickness for stars and gas). For each case, we derive the appropriate dispersion relation in the WKB approximation, and study the dynamical effect of the disc thickness on the life-time of spiral density waves via a parametric approach. We find the generic trend that the effective reduction in disc self-gravity due to disc thickness makes it more stable against the non-axisymmetric perturbations, and shortens the life-span of the spiral density waves. Further, the interstellar gas and the disc thickness are shown to have a mutually opposite dynamical effect on the disc stability as well as the longevity of the spiral density waves. While the gas supports the  non-axisymmetric features for a longer time, the disc thickness has an opposite, quenching effect. Consequently, the net change is set by the relative dominance of the opposite effects of the interstellar gas and the disc thickness.
 }


   \maketitle
%

\section{Introduction} 
Spiral features are one of the most common non-axisymmetric structures (apart from a bar) in disc galaxies in the local Universe \citep[e.g][]{Elmegreenetal2011,Yuetal2018,Savchenkoetal2020}. The occurrence of spiral structure in high-redshift (up to $z \sim$ 1.8) disc galaxies is also known observationally  \citep[e.g.][]{Elmegreenetal2014,Willetetal2017,Hodgeetal2019}. Simultaneous occurrence of spiral structure with an $m=2$ bar or an $m=1$ lopsidedness in disc galaxies is commonplace \citep[e.g., see][]{RixandZaritsky1995, Bournaudetal2005, Butaetal2010,Zaritskyetal2013,Kruketal2018,Ghoshetal2021}. The Milky Way is a barred spiral galaxy \citep{Weinberg1992, Gerhard2002} as well. In the past, a plethora of physical mechanisms ranging from bar-induced spirals \citep[e.g.,][]{Salo2010}, or due to tidal encounters \citep[e.g.,][]{Toomre1972, Dobbsetal2010}, or swing amplification of noise \citep{GoldreichLyden65, JulainToomre66, Toomre81}, due to disc response to giant molecular clouds \citep{Donghia2013}, due to interactions with other spirals \citep{Masset1997}, and due to recurrent groove modes \citep{SellwoodLin1989, Sellwood2012, SellwoodandCarlberg2019} to manifold-driven spirals \citep{Athanassoula2012}, have been proposed for exciting spirals in disc galaxies. 
\par
Regardless of the physical mechanism(s) triggering the spiral instability in the disc, it is vital to address the \textit{nature} and the longevity of spirals. Past studies of \citet{GoldreichLyden65, JulainToomre66, Toomre81} proposed the spiral arms as transient \textit{material arms} arising due to joint effects of epicyclic motion of stars, shear due to differential rotation, and the disc self-gravity. On the other hand, another set of studies by \citet{LinandShu1964,LinandShu1966} envisaged the spirals as \textit{quasi-stationary density waves} which rotate with respect to the disc with a well-defined pattern speed \citep[for a detailed recent review, see][]{Shu2016}. Implicit to the formalism of swing amplification, the material arms do not last for a long time (beyond a few dynamical timescales or a few $\times 10^8$ yr), and get wound up due to the disc differential rotation, whereas by definition, the spiral density waves are assumed to be stationary or they last forever.  However, \citet{Toomre1969} showed that, due to the radial group transport, a wavepacket made of such density waves eventually gets destroyed within $\sim 1 \Gyr$ time-scale, thereby posing a challenge to the \textit{stationary} picture of the density wave theory. Spirals, generated in $N$-body simulations, are almost always found to be a transient phenomenon \citep[e. g., see][ for a review the reader is referred to \citet{DobbsandBaba2014}]{Sellwood2011,Grandeetal2012,Babaetal2013}. However, a high-resolution $N$-body simulation by \citet{Donghia2013} showed that the spirals can sustain (at least in the statistical sense) for much longer time-scale due to the \textit{non-linear} disc response to the  perturbation caused by giant molecular cloud-like mass concentration. Another recent work by \citet{SahaandElmegreen2016} showed that bulges play a pivotal role in the sustenance of spirals in $N$-body simulations, and intermediate-sized bulges can help the spiral density wave to  last longer ($\sim 5 \Gyr$) by providing a high Toomre $Q$-barrier in the inner region. Interstellar gas is shown to play a pivotal role too in the longevity of spiral density waves in disc galaxies. Earlier work by  \citet{SellwoodCarlberg1984} demonstrated the role of a dissipative component in cooling the stellar disc and facilitating the generation of fresh spiral waves. \citet{Jog1992} included gas along with stars in the framework of swing amplification mechanism for a more realistic treatment which yielded broad stellar arms as observed. Further,  \citet{GhoshJog2015, GhoshJog2016} showed that the interstellar gas helps spiral density waves to survive for a longer time (several billion years).
\par
Most of the past analytical studies of spiral structure in disc galaxies treated the galactic disc as being \textit{infinitesimally-thin}, for simplicity \citep[but see][]{GoldreichLyden65}. This assumption is valid when the height of the disc is small as compared to the wavelength of the perturbation \citep[see discussions in][]{Toomre1964,BT08}. However, in reality, a galactic disc has a finite height. 
\citet{Lopezetal2014} showed that the Galactic disc flares substantially between Galactocentric radii $8 \kpc$ to $25 \kpc$ \citep[also see][]{Lietal2019} with the thin-disc component having a scale-height of $\sim 300 \pc$ in the Solar neighbourhood \citep[e. g., see][]{Juricetal2008}. This flaring of galactic disc (i.e., increment of scale-height) is a generic phenomenon \citep[e.g., see][]{degrijs1997,NarayanJog2002a,SarkarandJog2019,Garciaetal2021}. In addition, the existence of a thick-disc component is now well-established observationally in both, external galaxies, as well as the Milky Way \citep[e.g., see][]{Tsikoudi1979,Burstein1979,Yoachim2006,Comenronetal2011a,Comeron2011b,Comeronetal2018}.  Spirals play a pivotal dynamical role in disc dynamics by transporting angular momentum \citep{LyndenBellKalnajs1972}; by causing radial migration of stars without heating \citep{SellwoodBinney2002, Roskaretal2008, SchonrichBinney2009}, and  by exciting vertical breathing motions \citep{Debattista2014,Faureetal2014,Ghoshetal2020}. Thus, a more realistic study of the spiral structure in disc galaxies should take into account the finite thickness of the disc.
\par
Past theoretical studies have shown that the introduction of a finite thickness of a galactic disc results in a net reduction in the radial force in the mid-plane \citep[e.g., see][]{Toomre1964,JogSolomon1984,Jog2014}. This, in turn, facilitates the galactic disc to become more stable against the axisymmetric perturbations \citep{Toomre1964,JogSolomon1984}. \citet{JulainToomre66} demonstrated that the finite thickness of the stellar disc decreases the amplitude of the density transforms, in a local patch of the disc. Further, \citet{GhoshJog2018} showed the generic trend of suppressing the growth of the \textit{swing-amplified} spirals due to the introduction of a finite thickness for the disc. However, little is known about the plausible role of the disc finite thickness on the radial group transport and the longevity of spiral density wave.
\par
In this paper, we investigate the physical effect of the finite thickness of a galactic disc on the persistence of spiral density wave, both in absence and in presence of the interstellar gas. To achieve that, we first derive the appropriate dispersion relation in the WKB (Wentzel-Kramers-Brillouin) limit, for a collisionless stellar disc with a finite thickness and for a gravitationally-coupled two-component (stars plus gas) system with different thickness for the stellar and gas discs. Then, we systematically vary different input parameters, namely, the Toomre $Q$ parameter,  thickness of the disc, and the gas fraction, to check the dependence of the longevity of the spiral density wave, if any, on the finite thickness of the disc. 
 The rest of the paper is organised as follows: 
 Sect.~\ref{sec:formulation} provides the derivation of the relevant dispersion relations in the WKB limit whereas Sect.~\ref{sec:dimen_disp} presents the dimensionless form of the corresponding dispersion relations. Sect.~\ref{sec:results} gives our results covering the effect of finite thickness on the persistence of the spiral density waves, with and without the interstellar gas. Sects.~\ref{sec:discussion} and~\ref{sec:conclusion} contain discussion and  the main findings of this work, respectively.

\section {WKB dispersion relation for galactic disc with finite thickness}
\label{sec:formulation}

Here, we first derive the dispersion relation for a one-fluid disc, and then extend to a collisionless stellar disc with finite height (Sects.~\ref{sec:one_comp_formulation}--\ref{sec:one_comp_stellardisk}). Finally, we consider a more realistic gravitationally-coupled two-component (stars plus gas) system where the stellar and the gaseous discs can have different scale-heights (Sect.~\ref{sec:two_comp_formulation}), as observed in real galaxies. The underlying disc is taken to be axisymmetric and the spiral structure is treated as only a small perturbation on the steady-state axisymmetric disc, so that linear perturbation approach will be valid \citep[for details see][]{BT08}. A cylindrical coordinate system $(R, \phi, z)$ is used throughout the formulation.

\subsection {One-component fluid disc}
\label{sec:one_comp_formulation}

Following the treatment given in \citet{BT08}, we start with an {\it infinitesimally} thin fluid disc, and then we modify the formulation by introducing the effect due to the finite thickness of the fluid disc. The fluid disc is characterised by the disc surface density $\Sigma_{\rm d}$ and the sound speed $c$.  We assume that the pressure acts only in the disc plane.
Now, for such a system, the Euler's equations of motion in the cylindrical coordinates become

\begin{equation}
\frac{\partial v_R}{\partial t}+ v_R \frac{\partial v_R}{\partial R}+\frac{v_{\phi}}{R} \frac{\partial v_R}{\partial \phi} -\frac{v^2_{\phi}}{R} = - \left(\frac{\partial (\Phi+ {\mathcal H})}{\partial R}\right)_{z = 0}\,,
\label{eq:one}
\end {equation}
\noindent and,

\begin{equation}
\frac{\partial v_{\phi}}{\partial t}+ v_R \frac{\partial v_{\phi}}{\partial R}+\frac{v_{\phi}}{R} \frac{\partial v_{\phi}}{\partial \phi} +\frac{v_R v_{\phi}}{R}  = - \frac{1}{R} \left(\frac{\partial (\Phi+{\mathcal H})}{\partial \phi}\right)_{z = 0}\,.
\label{eq:two}
\end {equation}\
\noindent Here, ${\mathcal H}$ is the specific enthalpy of a polytropic fluid with an equation of state $p =K \Sigma_d^{\gamma}$ where $\gamma = (n+1)/n$, $n$ being the polytropic index, and $K$ being a proportionality constant, and the form of  ${\mathcal H}$ is given by \citep{BT08}
\begin{equation}
{\mathcal H} = \frac{\gamma}{\gamma-1} K \Sigma_{d}^{\gamma-1}\,.
\end{equation}
\noindent Now, assuming the spiral density wave to be a small perturbation, we write $v_R = v_{R_1}; v_{\phi} = v_{\phi_0} + v_{\phi_1}$, where $v_{R_1}$ and $v_{\phi_1}$ are small perturbations. Further assuming the random motion to be small compared to the rotation, from the Euler equation, it can be seen that the unperturbed motion gives rise to

\begin{equation}
v_{\phi_0} = \sqrt{R \frac{d \Phi_0}{dR}} = R \Omega (R)\,,
\end{equation}
\noindent where $\Phi_0$ is the unperturbed potential and $\Omega (R)$ is the circular frequency. The linear perturbed equations of motion become

\begin{equation}
\frac{\partial v_{R_1}}{\partial t}+ \Omega \frac{\partial v_{R_1}}{\partial \phi} -2\Omega(R)v_{\phi_1} = -\left(\frac{\partial (\Phi_1+{\mathcal H}_1)}{\partial R} \right)_{z=0}\,,
\label{eq:onef_pervr}
\end{equation}
\noindent and,

\begin{equation}
\frac{\partial v_{\phi_1}}{\partial t} +\left[\frac{d(\Omega R)}{dR}+\Omega\right]v_{R_1}+\Omega \frac{\partial v_{\phi_1}}{\partial \phi}= - \frac{1}{R}\left(\frac{\partial (\Phi_1+{\mathcal H}_1)}{\partial \phi 
} \right)_{z=0}\,,
\label{eq:onef_pervphi}
\end{equation}
\noindent where ${\mathcal H}_1$ is the perturbed specific enthalpy.

\subsubsection {Introduction of finite thickness of the fluid disc}
\label{sec:finite_thickness_onefluid}

For simplicity, we assume that the disc has a constant density that does not vary with $z$, and the disc has a total thickness of $2h$. 
For an infinitesimally thin, axisymmetric disc, and an axisymmetric perturbation, the solution of the Poisson equation is given by \citep{Toomre1964}

\begin{equation}
 \Phi_1 =  - (2 \pi G / |k|) \Sigma_1 exp(- k |z|) \,,
 \label{eq:potential_WKB}
\end{equation}
where $k$ is the wavenumber of perturbation and $\Phi_1$ and $\Sigma_1$ are the perturbation (or imposed) potential  and the corresponding surface density, respectively. The perturbation surface density is taken to have the form exp$[i(kr - \omega t)]$.
For such a disc, the radial force at the mid-plane ($z=0$) due to a vertical layer between $z$ and $z+dz$ is 
proportional to exp$[- k |z| dz]$. Hence, the net radial force at $z=0$  is obtained by integrating over $z$ to be \citep[ see][]{Toomre1964}

\begin{equation}
  \left(\frac{\partial  \Phi_1}{\partial R}\Big |_{z=0}\right)_{\rm net} = - i 2 \pi G {\Sigma_1}  \delta\,,
\end{equation}
\noindent where $\delta$ is the reduction factor which denotes the reduction in the radial force at the mid-plane due to the finite height for a constant density disc. This can also be thought of as a reduction in the disc surface density, and its form is given as \citep{Toomre1964}

\begin{equation}
\delta = [1- exp(- k h)]/kh \,.
\label{eq:reduc_fac}
\end{equation}
 In an analogous fashion, by integrating the contribution of force due to layers at different $z$, 
 we can show that the azimuthal force in the mid-plane is also reduced by an identical reduction factor.

We point out that, the perturbation potential and the surface density in the framework of WKB approximation, follow the same relation (Eq.~(\ref{eq:potential_WKB})) as above \citep[see Eqs. (6)-(18) in][]{BT08}. Hence,  a similar analysis (as mentioned above) will yield a same reduction factor $\delta$ (Eq.~(\ref{eq:reduc_fac})) in the radial, as well as the azimuthal force at mid-plane, for the WKB approximation as well. Therefore, in case of a fluid disc with a finite thickness, Eqs.~(\ref{eq:onef_pervr}) and ~(\ref{eq:onef_pervphi}) become

\begin{equation}
\frac{\partial v_{R_1}}{\partial t}+ \Omega \frac{\partial v_{R_1}}{\partial \phi} -2\Omega(R)v_{\phi_1} = -\left(\frac{\partial \Phi_1}{\partial R} \delta +  \frac{\partial {\mathcal H}_1}{\partial R}\right)_{z=0}\,,
\label{eq:onef_pervr_fh}
\end{equation}
\noindent and,

\begin{equation}
\frac{\partial v_{\phi_1}}{\partial t} +\left[\frac{d(\Omega R)}{dR}+\Omega\right]v_{R_1}+\Omega \frac{\partial v_{\phi_1}}{\partial \phi}= - \frac{1}{R}\left(\frac{\partial \Phi_1}{\partial \phi} \delta +  \frac{\partial {\mathcal H}_1}{\partial \phi}\right)_{z=0}\,.
\label{eq:onef_pervphi_fh}
\end{equation}

We assume the trial solutions are of the form

\begin{equation}
\begin{split}
v_{R_1} = Re \left[v_{R_a} (R) e^{i(m\phi + k R  - \omega t) }\right] \\
 v_{\phi_1} = Re \left[v_{\phi_a} (R) e^{i(m\phi  + k R  - \omega t) }\right]\\
\Phi_1 = Re \left[\Phi{_a} (R) e^{i(m\phi  + k R - \omega t)  - k |z|}\right]\\
 \Sigma_{d_1} = Re \left[\Sigma_{d_a} (R) e^{i(m\phi + k R  - \omega t) }\right]\\
 {\mathcal H}_1 = Re \left[{\mathcal H}_a (R) e^{i(m\phi + k R  - \omega t) }\right]\,.
 \label{eq:trial_sol}
\end{split}
\end{equation}
\indent Putting these in Eqs.~(\ref{eq:onef_pervr_fh}) and~(\ref{eq:onef_pervphi_fh}), and after some algebraic simplification we get,

\begin{equation}
v_{R_a} = \frac{i}{\Delta}\left[(\omega -m \Omega)\left(\delta \frac{d \Phi_a}{dR} + \frac {d {\mathcal H}_a}{dR}\right) - \frac{2m \Omega}{R}(\Phi_a \delta + {\mathcal H}_a)\right]\,,
\label{eq:per_vra_onef}
\end{equation}
\noindent and,

\begin{equation}
v_{\phi_a} = -\frac{1}{\Delta} \left[2B \left(\delta \frac{d \Phi_a}{dR}  + \frac {d  {\mathcal H}_a}{dR}\right) + \frac{m(\omega-m\Omega)}{R}\left(\Phi_a \delta + {\mathcal H}_a\right)\right]\,,
\label{eq:per_vphia_onef}
\end{equation}
\noindent where $B$ is the Oort constant and $\Delta \equiv \kappa^2 - (\omega-m\Omega)^2$, $\kappa$ being the epicyclic frequency.  Also, the perturbed equation of state gives ${\mathcal H}_a = c^2 \frac{\Sigma_{d_a}}{\Sigma_0}$, where $\Sigma_0$ is the unperturbed disc surface density \citep[for details see][]{BT08}.

Similarly, the perturbed continuity equation in cylindrical coordinates is 

\begin{equation}
\frac{\partial \Sigma_{d_1}}{\partial t}+\Omega \frac{\partial \Sigma_{d_1}}{\partial \phi}+\frac{1}{R}\frac{\partial}{\partial R}\left(R v_{R_1} \Sigma_0\right) + \frac{\Sigma_0}{R}\frac{\partial v_{\phi_1}}{\partial \phi} = 0\,,
\end{equation}
\noindent which, after substituting the trial solution (Eq.~(\ref{eq:trial_sol})) becomes

\begin{equation}
-i(\omega-m\Omega)\Sigma_{d_a} e^{i kR} +\frac{1}{R}\frac{d}{dR}\left(R v_{R_a} e^{i kR} \Sigma_0\right)+\frac{im\Sigma_0}{R}v_{\phi_a} e^{ i kR} =0\,.
\label{eq:per_eq_cont_onef}
\end{equation}

\subsubsection {Dispersion relation}
\label{sec:dispreln_onef}

Now, we invoke the WKB approximation to derive an analytical dispersion relation. In this limit, Eqs.~(\ref{eq:per_vra_onef})-(\ref{eq:per_vphia_onef}) reduce to \citep[for details, see][]{BT08}
\begin{equation}
\begin{split}
v_{R_a} = -\frac{(\omega-m\Omega)}{\Delta}k(\Phi_a \delta+{\mathcal H}_a) \hspace{0.2 cm }\mbox{and} \hspace{0.2 cm }
 v_{\phi_a} = - \frac{2iB}{\Delta}k(\Phi_a \delta+{\mathcal H}_a)\,,
 \end{split}
 \label{eq:meanVel_WKB}
\end{equation}
\noindent Also, the continuity equation (Eq.~(\ref{eq:per_eq_cont_onef})) reduces to 

\begin{equation}
-(\omega-m\Omega)\Sigma_{d_a}+k \Sigma_0 v_{R_a} =0\,,
\label{eq:per_cont_eqn_onf_final}
\end{equation}
\noindent as in the WKB approximation,  the last term in Eq.~(\ref{eq:per_eq_cont_onef})  involving $v_{\phi_a}$ is smaller than the first two terms and can be dropped \citep{BT08}.
Now, on substituting the values for $v_{R_a}$, ${\mathcal H}_a$  as obtained above and $\Phi_a = -2\pi G \Sigma_{a} /|k|$, as obtained in the WKB limit using Eqs.~(\ref{eq:potential_WKB}) and~(\ref{eq:trial_sol}), where $\Phi_a$  and $\Sigma_a$ are the imposed perturbation potential and the corresponding imposed perturbation surface density; and after some algebraic manipulation, Eq.~(\ref{eq:per_cont_eqn_onf_final})  reduces to

\begin{equation}
\Sigma_{d_a} = \frac{2 \pi G |k| \Sigma_0 \delta}{\kappa^2 - (\omega-m\Omega)^2+c^2 k^2} \Sigma_a\,,
\end{equation}
\noindent where $\Sigma_{d_a}$ is the disc response density. Now, for a {\it self-sustained} density wave, the quantities 
$\Sigma_a$ and $\Sigma_{d_a}$
should be equal, and hence the dispersion relation becomes \citep{BT08}

\begin{equation}
(\omega-m\Omega)^2 = \kappa^2- 2 \pi G |k| \Sigma_0 \delta + c^2 k^2\,,
\label{eq:disp_reln_onef_hf}
\end{equation}
\noindent where $m$ is a positive integer, denoting the $m$-fold rotational symmetry of the perturbation.
We note that, in the limit of $h \rightarrow 0$, so that $\delta \rightarrow 1$, the dispersion relation (Eq.~(\ref{eq:disp_reln_onef_hf})) reduces to the corresponding dispersion relation for an infinitesimally-thin fluid disc \citep[see][]{BT08}, as expected. In other words, Eq.~(\ref{eq:disp_reln_onef_hf}) is a generalisation of the dispersion relation for an infinitesimally thin fluid disc. Here, the additional factor $\delta$ comes in from the inclusion of the effect of finite height in the calculation.

\subsection {One-component stellar disc}
\label{sec:one_comp_stellardisk}

 A {\it cold} stellar disc is dynamically equivalent to a fluid disc with zero pressure or zero enthalpy, and hence, the perturbation $\bar v_{R_1}$ in the mean radial velocity of the stars can be obtained from Eqs.~(\ref{eq:onef_pervr}) and (\ref{eq:onef_pervphi}) (assuming zero pressure) as \citep{BT08}
\begin{equation}
\bar v_{R_a} = -\frac{\omega-m\Omega}{\Delta} k \Phi_a\,.
\end{equation}
\noindent where $\Delta$ is as defined earlier.
If we assume the disc to have a finite thickness of $2h$ and a constant density $\rho$ (as in Sect.~\ref{sec:finite_thickness_onefluid}), then in an analogous fashion, the radial force in the mid-plane is reduced by a reduction factor $\delta$ (defined in Eq.~(\ref{eq:reduc_fac})). Then following the same procedure as done in Sect.~\ref{sec:finite_thickness_onefluid}, namely solving the Euler equation (Eq.~\ref{eq:onef_pervr_fh})  with 
zero enthalpy, the solution for $\bar v_{R_a}$ is obtained in the WKB limit to be

\begin{equation}
\bar v_{R_a} = -\frac{\omega-m\Omega}{\Delta} k (\Phi_a\, \delta)\,.
\label{eq:perVra_onecomp_fh}
\end{equation}
\noindent As expected, this can be obtained from Eq.~(\ref{eq:meanVel_WKB}) by setting the enthalpy to be zero.

 Now, if the stellar disc is not sufficiently  {\it cold}, i.e., if the typical epicyclic amplitude is not small enough as compared to the wavelength ($2 \pi/k$) of perturbation, then the net response velocity measured at a given location is due to stars with large epicyclic amplitudes and hence would have sampled different values of the spiral potential. This results in a partial cancellation in the mean velocity perturbation response to the imposed potential \citep[for details see][]{BT08}. Therefore, the resulting expression for the mean velocity perturbation for a thin disc is given by
 
 \begin{equation}
\bar v_{R_a} = -\frac{\omega-m\Omega}{\Delta} k \Phi_a \mathcal{F}\,,
\end{equation}
\noindent where $\mathcal{F}$ is the factor by which the response of the disc to a given spiral perturbation is diminished below the value for a cold disc. The form of $\mathcal{F}$ is discussed later in Sect.~\ref{sec:dimenls_onecomp_stars_fh}.

Now, it is straightforward to show that, for a stellar disc with finite thickness, where we start with $\bar v_{R_a}$ as for a finite hight case above (Eq.~(\ref{eq:perVra_onecomp_fh})), and then taking account of the reduction in the collisionless disc response when the disc is not cold, then these two effects  manifest simultaneously and the  expression for the mean velocity perturbation becomes

\begin{equation}
\bar v_{R_a} = -\frac{\omega-m\Omega}{\Delta} k (\Phi_a \delta) \mathcal{F} \,.
\label{eq:vRa_onecomp_hf_final}
\end{equation}
\noindent Also, the perturbed continuity equation in the WKB limit becomes \citep[see e.g.,][]{BT08}
\begin{equation}
-(\omega-m\Omega)\Sigma_{d_a}+k \Sigma_0 \bar v_{R_a} =0\,.
\label{eq:pertb_jeans_onf_fh}
\end{equation}
\noindent Note that this is identical to Eq.~(\ref{eq:per_cont_eqn_onf_final}) since the continuity equation has the same form for 
fluid and collisionless cases. On substituting $\bar v_{R_a}$ from Eq.~(\ref{eq:vRa_onecomp_hf_final}), the above reduces to

\begin{equation}
-(\omega-m\Omega)\Sigma_{d_a}+k \Sigma_0 \left[-\frac{\omega-m\Omega}{\Delta} k (\Phi_a \delta) \mathcal{F} \right] =0\,.
\label{eq:vRa}
\end{equation}

Finally, invoking the WKB approximation as before (Sect.~\ref{sec:finite_thickness_onefluid}), 
we substitute $\Phi_a = -2\pi G \Sigma_a/|k|$ (Sect.~\ref{sec:one_comp_stellardisk}), and obtain 
\begin{equation}
\Sigma_{d_a} = \frac{2 \pi G \Sigma_0|k|}{\Delta} \delta {\mathcal{F}} \Sigma_a\,.
\end{equation}

Again, for a self-sustained density wave, the disc response surface density ($\Sigma_{d_a}$) should be equal to the imposed surface density ($\Sigma_a$). This, in turn, yields the dispersion relation for the collisionless disc in the WKB limit as 

\begin{equation}
(\omega-m\Omega)^2 = \kappa^2 -2 \pi G |k| \Sigma_0 \delta \mathcal{F}\,.
\label{eq:disp_reln_stellardisk}
\end{equation}
\noindent Here also we note that in the limit of $h \rightarrow 0$, so that $\delta \rightarrow 1$, the dispersion relation (Eq.~(\ref{eq:disp_reln_stellardisk})) reduces to the corresponding dispersion relation for an infinitesimally-thin collisionless stellar disc \citep[see][]{BT08}, as expected. In other words, Eq.~(\ref{eq:disp_reln_stellardisk}) is a generalisation of the dispersion relation for an infinitesimally-thin stellar disc. Here, the additional factor $\delta$ comes in from the inclusion of the effect of finite height in the calculation.

\subsection {Two-component star-gas system with different finite thickness}
\label{sec:two_comp_formulation}

Here, we treat a galactic disc as a gravitationally coupled two-component (stars plus gas) system \footnote{ The main difference between this work with that presented in \citet{GhoshJog2015} is that here we take into account the finite thickness of both the stellar and the gaseous discs.}. The stellar disc is taken to be collisionless in nature and is characterised by a surface density $\Sigma_{\rm 0s}$, a one-dimensional velocity dispersion, $\sigma_{\rm s}$, and a total thickness of $2h_s$, while the  gas disc is treated as a fluid and is characterised by the surface density $\Sigma_{\rm 0g}$, a one-dimensional velocity dispersion or the sound speed, $c_{\rm g}$, and a total thickness of $2h_g$. Since the stars and the gas are gravitationally-coupled, therefore, their motion will be governed by the joint potential ($\Phi_{\rm tot}$) which is set by both the stellar and the gas discs, i.e., $\Phi_{\rm tot} = \Phi_{\rm s}+ \Phi_{\rm g}$. The right hand side of Eqs.~(\ref{eq:one}) and (\ref{eq:two}) will now contain a derivative of the total potential. Thus, the steady-state unperturbed motion will now be given by  

\begin{equation}
v_{\phi_{0i}} = \sqrt{R \frac{d \Phi_{\rm tot,0}}{dR}} = R \Omega(R)\,.
\end {equation}

Now, for such a system, the reduction due to the finite height affects the corresponding radial force of each component separately, and each component is affected by the net force due to both components \citep[see e.g.][]{JogSolomon1984}. Hence, the R.H.S. of the perturbed Euler equation (Eq.~(\ref{eq:onef_pervr_fh})) becomes $(\partial \Phi{s_1}/\partial R) \delta_s + (\partial \Phi_{g_1}/ \partial R) \delta_g + (\partial {\mathcal H}_{1_g}/\partial R $) for $i=g$ (gas), while for the collisionless stellar disc ($i=s$), only the first two terms above (without the enthalpy term) are kept, as discussed in Sect.~\ref{sec:one_comp_stellardisk}. The same correction applies to the azimuthal perturbed equation of motion as well. Now assuming trial solutions similar to those as shown in Eq.~(\ref{eq:trial_sol}), and invoking the WKB approximation, the solutions of the Euler equations give the amplitudes of perturbed velocity along the radial and the azimuthal directions for the stellar and the gas disc respectively as

\begin{equation}
\begin{split}
v_{R_{a_s}} = -\frac{(\omega-m\Omega)}{\Delta}k(\Phi_{a_s} \delta_1  {\mathcal{F}}+ \Phi_{a_g} \delta_g)\\
 v_{\phi_{a_s}} = - \frac{2iB}{\Delta}k(  \Phi_{a_s} \delta_1 {\mathcal{F}} + \Phi_{a_g} \delta_g)\,,
\end{split}
\label{eq:expre_vRa_twocomp_star}
\end{equation}
\noindent and,

\begin{equation}
\begin{split}
v_{R_{a_g}} = -\frac{(\omega-m\Omega)}{\Delta}k(\Phi_{a_s} \delta_s+ \Phi_{a_g} \delta_g +   {\mathcal H}_{a_g})\\
v_{\phi_{a_g}} = - \frac{2iB}{\Delta}k(\Phi_{a_s} \delta_s + \Phi_{a_g} \delta_g +{\mathcal H}_{a_g})\,.
\end{split}
\label{eq:expre_vRa_twocomp_gas}
\end{equation}
Substituting the expression for $v_{R_{a_s}}$ from Eq.~(\ref{eq:expre_vRa_twocomp_star}) in the perturbed continuity equation (analog of Eq.~(\ref{eq:pertb_jeans_onf_fh})), we get

\begin{equation}
\Sigma_{d_{a_s}}\left[{\kappa^2 - (\omega-m\Omega)^2} \right] = k \Sigma_{0_s} \left[ k ( \Phi_{a_s} \delta_1 {\mathcal{F}} + \Phi_{a_g} \delta_g) \right]\,.
\label{eq:resp_star_per}
\end {equation}

On substituting the expression for the perturbation potentials in terms of the surface densities in the WKB limit, namely $\Phi_{a_s} = - 2 \pi G \Sigma_{a_s}/|k|$ and similarly for $\Phi_{a_g}$, and setting $\Sigma_{a_s}$, the imposed or perturbation stellar surface density equal to $\Sigma_{d_{a_s}}$, the stellar response surface density for a self-consistent solution, the above reduces to 

\begin{equation}
\frac{\Sigma_{d_{a_s}}}{\Sigma_{a_g}} =  \frac {2 \pi G k \Sigma_{0_s} \delta_g}{\left(\kappa^2 - (\omega-m\Omega)^2 - 2 \pi G |k| \Sigma_{0_s} \delta_s {\mathcal{F}} \right)}\,.
\label{eq:meta_disp_star}
\end{equation}

Similarly for the gas disc, the continuity equation (Eq.~(\ref{eq:per_cont_eqn_onf_final})) combined with the perturbed velocity component in the WKB limit (Eq.~(\ref{eq:expre_vRa_twocomp_gas}))  gives:

\begin{equation}
\Sigma_{d_{a_g}}\left[{\kappa^2 - (\omega-m\Omega)^2} \right] = k \Sigma_{0_g} \left [ k ( \Phi_{a_s} \delta_s  + \Phi_{a_g} \delta_g+ {\mathcal H}_{a_g}) \right]\,.
\end {equation}
\noindent On substituting the expression for the perturbation potentials in terms of the surface densities in the WKB limit, namely $\Phi_{a_s} = - 2 \pi G \Sigma_{a_s}/|k|$ and similarly for $\Phi_{a_g}$; and further for self-consistency setting $\Sigma_{a_g} = \Sigma_{d_{a_g}}$, the above equation reduces to

\begin{equation}
\frac{\Sigma_{a_s}}{\Sigma_{d_{a_g}}} =  \frac {\left(\kappa^2 - (\omega-m\Omega)^2 +k^2 c_g^2 - 2 \pi G |k| \Sigma_{0_g} \delta_g \right)} {2 \pi G k \Sigma_{0_g} \delta_s}\,.
\label{eq:meta_disp_gas}
\end{equation}

Combining Eqs.~(\ref{eq:meta_disp_star}) and (\ref{eq:meta_disp_gas}) and setting the condition of self-consistency ($\Sigma_{a_s} = \Sigma_{d_{a_s}}$ and similarly for the gas disc), we get the dispersion relation for this joint star-gas system as

\begin{equation}
\frac{2\pi G \Sigma_{\rm 0s} |k|\delta_s{\mathcal F}\Big(\frac{\omega-m \Omega}{\kappa},\frac{k^2\sigma^2_{\rm s}}{\kappa^2}\Big)}{\kappa^2-(\omega-m\Omega)^2}+\frac{2\pi G \Sigma_{\rm 0g} \delta_g|k|}{\kappa^2-(\omega-m\Omega)^2+c^2_{\rm g} k^2} = 1\,.
\label{eq:dispreln_semi_twocomp}
\end{equation}

We checked that for an infinitesimally thin disc (where the reduction factors $\delta_i \rightarrow 1 \  (i= s, g)$, see Eq.~(\ref{eq:reduc_fac})), the above reduces to the dispersion relation for a star-gas case obtained in \citet{GhoshJog2015}, as expected.

We next define
\begin{equation}
\begin{split}
\alpha_{\rm s} = \kappa^2-2\pi G\Sigma_{\rm 0s}\delta_s|k| {\mathcal F}\Big(\frac{\omega-m \Omega}{\kappa},\frac{k^2\sigma^2_{\rm s}}{\kappa^2}\Big)\\
\alpha_{\rm g} = \kappa^2-2\pi G\Sigma_{\rm 0g}\delta_g|k|+k^2{c}^2_{\rm g}\\
\beta_{\rm s} = 2\pi G\Sigma_{\rm 0s}\delta_s|k|{\mathcal F}\Big(\frac{\omega-m \Omega}{\kappa},\frac{k^2\sigma^2_{\rm s}}{\kappa^2}\Big)\\
\beta_{\rm g} = 2\pi G\Sigma_{\rm 0g}\delta_g|k| \,.
\end{split}
\label{eq:paramteric_2comp}
\end{equation}

Upon substitution in Eq.~(\ref{eq:dispreln_semi_twocomp}) and after some algebraic simplification, we get

\begin{equation}
(\omega-m\Omega)^4-(\alpha_{\rm s}+\alpha_{\rm g})(\omega-m\Omega)^2+(\alpha_{\rm s}\alpha_{\rm g}-\beta_{\rm s}\beta_{\rm g})=0\,.
\end{equation}
This is a quadratic equation in $(\omega-m\Omega)^2$. Solving it we get
\begin{equation}
(\omega-m\Omega)^2=\frac{1}{2}\left[(\alpha_{\rm s}+\alpha_{\rm g})\pm \left\{(\alpha_{\rm s}+\alpha_{\rm g})^2-4(\alpha_{\rm s}\alpha_{\rm g}-\beta_{\rm s}\beta_{\rm g})\right\}^{1/2}\right]\,.
\end{equation}
The  additive root for $(\omega-m\Omega)^2$ always leads to a positive quantity, hence it indicates always oscillatory perturbations under all conditions \citep[same as for axisymmetric case; see][]{JogSolomon1984}. In order to study the stability of the system and its further consequences, we therefore consider only the negative root which is

\begin{equation}
(\omega-m\Omega)^2=\frac{1}{2}\left[(\alpha_{\rm s}+\alpha_{\rm g})- \left\{(\alpha_{\rm s}+\alpha_{\rm g})^2-4(\alpha_{\rm s}\alpha_{\rm g}-\beta_{\rm s}\beta_{\rm g})\right\}^{1/2}\right]\,.
\label{disp-main}
\end{equation}

We mention that, the underlying formalism, presented here, closely follows that presented in \citet{JogSolomon1984}.
This builds on the treatment for one-component stellar and gas discs of finite height as in Sects. 2.1 and 2.2. On solving the coupled equations, the resulting dispersion relation is given by Eq.~(\ref{eq:dispreln_semi_twocomp}), which has the solution Eq.~(\ref{disp-main}). Also, we caution the reader that the form for the solution of the dispersion relation (Eq.~(\ref{disp-main})) and the subsidiary variables $\alpha_{\rm s}$, $\alpha_{\rm g}$, $\beta_{\rm s}$ and $\beta_{\rm g}$ (Eq.~(\ref{eq:paramteric_2comp})) (in terms of which Eq.~(\ref{disp-main}) is written) may appear similar 
to that in \citet[which treated a two-fluid case]{JogSolomon1984}, and \citet[which treated a star-gas case where the effect of stellar dispersion was included in terms of the reduction factor $\mathcal{F}$]{GhoshJog2015}. In fact these were defined to have a similar form by construction because all three are two-component formulations and have a similar underlying mathematical symmetry, but with different treatment for stars and gas, and the current work includes the effect of finite height.

\section {Dimensionless form of dispersion relations in the WKB limit}
\label{sec:dimen_disp}
In the literature, usually the dispersion relations are described in terms of some dimensionless quantities, for the sake of convenience. Here, we follow the same procedure for the dispersion relations derived in the previous sections.

\subsection{One-component fluid disc with finite thickness}
\label{sec:One-component fluid disc with finite thickness}

Dividing both sides of Eq.~(\ref{eq:disp_reln_onef_hf}) by $\kappa^2$, and after some algebraic simplification  we get
\begin{equation}
s^2 = 1 - x \delta+ \frac{1}{4} x^2 Q^2\,,
\end{equation}
\noindent where $s=({\omega-m\Omega})/{\kappa}$, and and $x$ (= $|k|/k_{\rm crit}$) are the dimensionless frequency and wavenumber of the perturbation, respectively; and $k_{\rm crit} (= \kappa^2 / 2 \pi G  \Sigma_{\rm 0})$  is the largest stable wavenumber for a pressure-less stellar disc. $Q (=\kappa c/\pi G \Sigma_0)$ is the usual Toomre $Q$ parameter for a fluid disc \citep{Toomre1964}. Also, $|k|h$ can be expressed as $|k|h = (|k|/k_{crit}) \times (k_{crit}h) = x\beta$. Here, $\beta$ is defined to be equal to $k_{\rm crit}\times h$. Therefore, the dispersion relation becomes

\begin{equation}
s^2 = 1 - x \delta+ \frac{1}{4} x^2 Q^2\,,
\end{equation}
\noindent where the form of $\delta$ reduces to

\begin{equation}
\delta = \frac{1-exp(-x\beta)}{x\beta}\,.
\label{eq:dimen_delta}
\end{equation}
Now, the value of $\beta$ is dependent on the chosen values of $k_{\rm crit}$, i.e., for the same thickness of a disc, the values of $\beta$ will be different depending on the values of $k_{\rm crit}$. We note that in the Solar neighbourhood, a circular velocity ($v_{\rm c}$) of $\sim$ 220 km s$^{-1}$ and $\Sigma$ $\sim$ 45 M$_{\odot}$ pc$^{-2}$ \citep[e.g. see][]{Meraetal1998,NarayanJog2002} will produce $k_{\rm crit}$ $\sim$ 1 kpc$^{-1}$. However, recent studies have reported slightly different values for the circular velocity and the Solar position \citep[e.g., see][]{Gillessenetal2009,Schoenrichetal2010,Schoenrich2012,McMillanetal2018,Schoenrichetal2019}. The Galactocentric distance of the Sun is $8.27 \kpc$ \citep{Schoenrich2012} which is in agreement with the other measurements \citep[within their error-bars][]{Gillessenetal2009,McMillanetal2018}. Also, the circular velocity at the Solar radius is $237.8 \kms$ \citep{Schoenrichetal2010} which, in turn, gives circular frequency at the Solar radius as $28.8 \kmsk$. Assuming a flat rotation curve, the corresponding epicyclic frequency becomes $40.7  \kmsk$. We use these latest values of $\Omega$ and $\kappa$ and consider the group transport at the solar neighbourhood ($R = 8.27$ kpc) in the subsequent sections, unless stated otherwise. Using these recent values, we estimate $k_{\rm crit}$ $\sim$ 1.3 kpc$^{-1}$. Therefore, for the sake of uniformity, we chose $k_{\rm crit}$ = 1.3 kpc$^{-1}$ for all cases considered here, unless stated otherwise. The resulting behaviour of the reduction factor $\delta$ is shown in Appendix~\ref{appen:reduc}. 
\subsection{One-component stellar disc with finite thickness}
\label{sec:dimenls_onecomp_stars_fh}

Dividing both sides of Eq.~(\ref{eq:disp_reln_stellardisk}) by $\kappa^2$ and after some algebraic simplification we get

\begin{equation}
s^2 = 1 - x \delta \mathcal{F}(s, \chi)\,,
\label{eq:disprel_onecomp_final}
\end{equation}
\noindent where, $\chi$= ${k^2\sigma^2_{\rm s}}/{\kappa^2}$ = $0.286 Q_{\rm s}^2 x^2$, and $\delta$ is already given in Eq.~(\ref{eq:dimen_delta}). The form for ${\mathcal F}$ for a razor-thin disc whose stellar equilibrium state is described by the Schwarzschild distribution function, is given by \citep{BT08}: 
\begin{equation}
{\mathcal F}(s, \chi)=\frac{2}{\chi}\exp(-\chi)(1-s^2)\sum_{n=1}^\infty\frac{I_n(\chi)}{1- s^2/n^2}\,,
\end{equation}
\noindent where $I_n$ is the modified Bessel function of first kind.

\subsection {Two-component star-gas system with different finite thickness}

Proceeding as before, on dividing both sides of Eq.~(\ref{disp-main}) by $\kappa^2$ and after some algebraic simplification we get

\begin{equation}
s^2=\frac{1}{2}\left[(\alpha'_{\rm s}+\alpha'_{\rm g})-\left \{(\alpha'_{\rm s}+\alpha'_{\rm g})^2-4(\alpha'_{\rm s}\alpha'_{\rm g}-\beta'_{\rm s}\beta'_{\rm g})\right \}^{1/2}\right]\\,
\label{disp-twocom}
\end{equation}
where
\begin{equation}
\begin{split}
\alpha'_{\rm s}=1-(1-\epsilon)x\delta_s{\mathcal F}(s,\xi)\\
\alpha'_{\rm g}=1-\epsilon x\delta_g+\frac{1}{4}Q^2_{\rm g}\epsilon^2x^2\\
\beta'_{\rm s}=(1-\epsilon)x\delta_s{\mathcal F}(s,\xi) \\
\beta'_{\rm g} = \epsilon x\delta_g\\
\end{split}
\end{equation}
where, $\xi$= ${k^2\sigma^2_{\rm s}}/{\kappa^2}$ = $0.286 Q_{\rm s}^2 (1-\epsilon)^2 x^2$.  The three dimensionless parameters $Q_{\rm s}$, $Q_{\rm g}$ and $\epsilon$ are, respectively the Toomre $Q$ parameters for stars $Q_{\rm s}$(=$\kappa \sigma_{\rm s} /(3.36 G \Sigma_{\rm 0s})$), and for gas $Q_{\rm g}$ = ($\kappa c_{\rm g} /(\pi G \Sigma_{\rm 0g})$)
and,  $\epsilon$ =${\Sigma_{\rm 0g}}/( {\Sigma_{\rm 0s}+\Sigma_{\rm 0g}})$ the gas mass fraction in the disc, respectively. Also, the forms of $\delta_i$ are given by

\begin{equation}
\delta_i = \frac{1-exp(-x\beta_i)}{x\beta_i}\,.
\end{equation}
\noindent where $i =s, g$ for stars and gas, respectively. We note that, when the height $h_i \rightarrow 0$, and so $\delta_i \rightarrow 1$,  Eq.~(\ref{disp-twocom}) reduces to the dispersion relation for the gravitationally coupled two-component (stars plus gas) system which is infinitesimally thin \citep[see][]{GhoshJog2015}, as expected.

\section{Results}
\label{sec:results}
Here, we present the results related to the dynamical effect of inclusion of a finite thickness of the disc, with or without the presence of the interstellar gas on the disc stability and the longevity of a spiral density wave. For a bi-symmetric ($m=2$) spiral density wave, the pattern speed $\Omega_{\rm p} = \omega/2$ and $s = (\omega - 2\Omega )/ \kappa  = 2 (\Omega_{\rm p} - \Omega)/ \kappa$. Here, $s=0$ corresponds to the corotation (CR) point and $|s|$ = 1 gives the Lindblad resonances \citep{GhoshJog2016}.  %
\par
To quantify the disc stability, we define a quantity $|s|_{\rm cut-off}$ as the lowest value of the dimensionless frequency ($s$), corresponding to an appropriate dispersion relation, for which one is able to obtain a real or stable wave solution at any given $R$ \citep[for details, see e.g.,][]{BT08,GhoshJog2016}.  For $s$ lying in the forbidden region namely, between the corotation or $s=0$ and $s_{\rm cut-off}$, the solution is imaginary and the wave is transient, at any given radius $R$ \citep[for details see discussions in][]{BT08,GhoshJog2016}. In other words, the value of $|s|_{\rm cut-off}$ denotes the edge of the forbidden region. A decrease in the value of $|s|_{\rm cut-off}$ signifies the decrease in the forbidden zone; this results in the disc becoming more prone to being unstable against the perturbations. Conversely, an increase in the value of $|s|_{\rm cut-off}$ denotes the increase in the forbidden zone, and the disc becomes more stable against the perturbations.
\par
As for the longevity of the spiral density wave, following \citet{Toomre1969}, we study the radial group transport of a wavepacket of such a density wave. To achieve that, we calculate the radial group velocity from the local dispersion relation, appropriate for a particular system considered here. It is  known that the information from a disturbance, generated at a certain radius $R$, propagates in the disc with its group velocity $v_{\rm g}$. When the medium is inhomogeneous, the group velocity, at a given radius $R$, is defined as \citep[e. g., see][]{Whitham1960,Lighthill1965} 

\begin{equation}
v_g (R) = \frac{\partial \omega (k, R)}{\partial k}\,.
\label{eq:grp_velocity}
\end{equation}
\noindent The value of the group velocity ($v_{\rm g}$), at a given $R$, can be estimated from the slope of the local dispersion relation (when expressed in a dimensionless form) by using the following equation \citep[e. g., see][]{Toomre1969,BT08}

\begin{equation}
v_g (R) = sgn(ks) \left(\frac{\kappa}{k_{\rm crit}}\right)  \frac{ds}{dx}\,.
\label{eq:grp_velocity_mod}
\end{equation}
\noindent Here, $s$ and $x$ are the dimensionless frequency and the wavenumber of the perturbation, respectively, and $sgn (ks) = \pm 1$ depending on whether $ks >0$ or $ks <0$. Here, the slope is obtained at a point $x$ where the observed value of $s$ intersects the dispersion relation curve. Thus, the location (or the $x$ value) where the slope is to be calculated is determined by the pattern speed ($\Omega_{\rm p}$) value as well as the underlying mass distribution (which in turn sets the values of $\Omega$ and $\kappa$), for more details see Sect.~\ref{sec:results_onecomp}.
\par
 A decrease in the group velocity implies that a wavepacket of such density wave would take a longer time to reach the centre of the disc, and eventually get absorbed. In other words, a decrease in the group velocity signifies a longer persistence of the spiral density waves in the disc \citep[for further discussions, see, e. g., ][]{Toomre1969,GhoshJog2015}. Sect.~\ref{sec:results_onecomp} provides the details of the effect of finite thickness on the disc instability and longevity of spiral density wave for the one-component stellar disc whereas Sect.~\ref{sec:results_twocomp} gives the same for the gravitationally-coupled two-component (stars plus gas) disc. We mention that the methodology of investigating the disc stability against the $m=2$ spiral density wave is similar to that in \citet{GhoshJog2016} whereas the treatment for calculation of the group velocity is similar to that in \citet{GhoshJog2015}, except here we use the appropriate dispersion relations for a one-component stellar disc, and a two-component disc with \textit{finite height}, as derived in the previous section.

\subsection{Effect of finite thickness on one-component stellar disc}
\label{sec:results_onecomp}

\begin{figure}
\centering
    \includegraphics[width=1.03\linewidth]{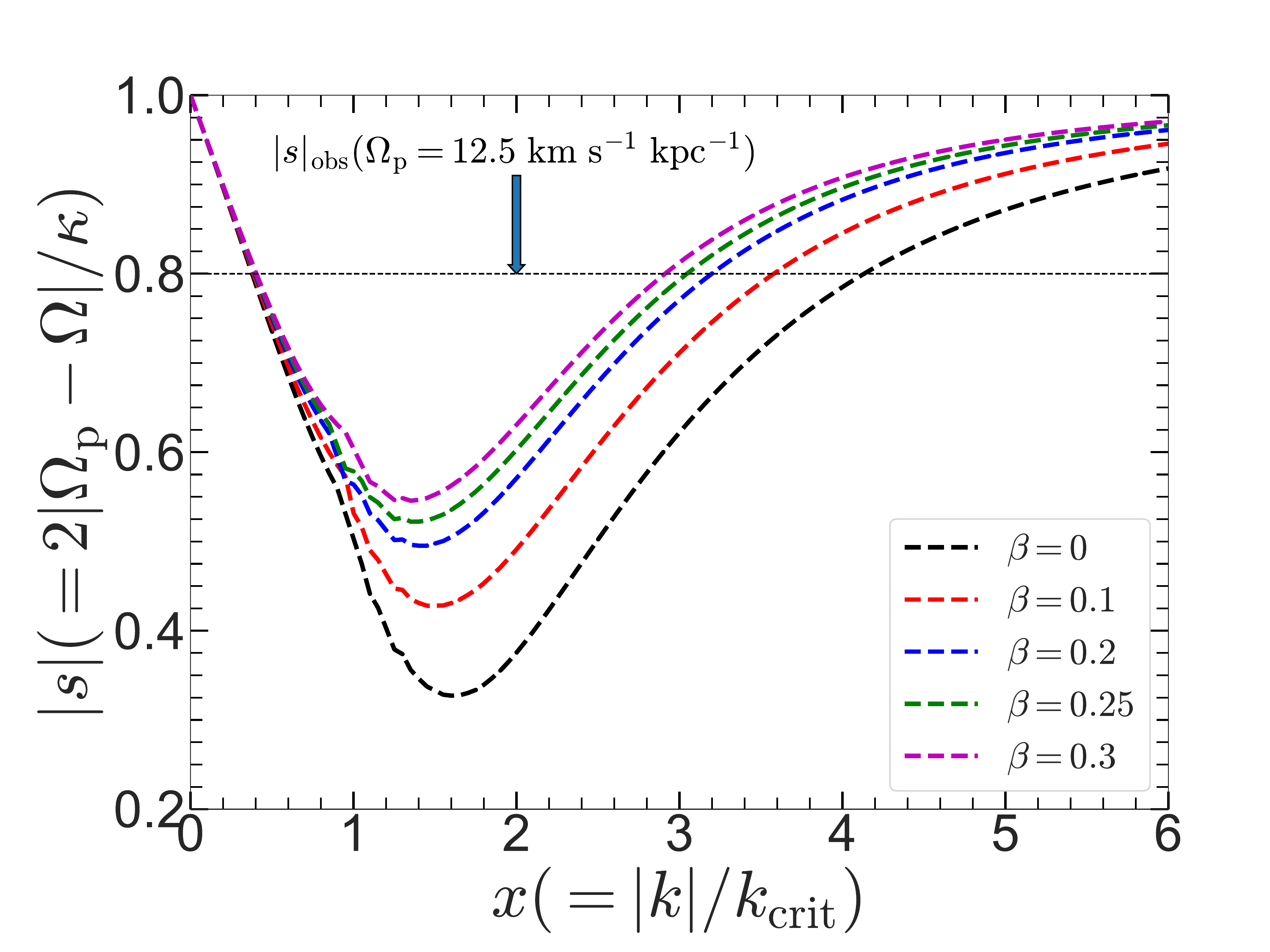}
\caption{Dispersion relation for the one-component stellar disc (Eq~(\ref{eq:disprel_onecomp_final})) is shown for $Q =1.1$ and for $\beta$ = 0.1--0.3. For comparison, the corresponding dispersion relation for the zero-thickness (or infinitesimally-thin) stellar disc is also shown (black dashed line). The horizontal dashed line (in black) denotes the $|s|_{\rm obs}$ value (calculated at $R = 8.27 \kpc$) corresponding to an assumed pattern speed of $12.5 \kmsk$.}
\label{fig:dispreln_onecomp}
\end{figure}

 To study the dynamical effect of the disc finite thickness, first we calculate the dispersion relation (Eq.~(\ref{eq:disprel_onecomp_final})) for Toomre $Q = 1.1$ while varying the disc thickness, or equivalently $\beta$ from 0.1 to 0.3. Fig.~\ref{fig:dispreln_onecomp} shows the corresponding dispersion relations. For comparison, we also show the corresponding dispersion relation for an infinitesimally-thin stellar disc with $Q =1.1$. \citet{Toomre1969} assumed an $Q =1$ which denotes the neutral stability of the stellar disc. Here, instead, we assume a slightly higher value of $Q =1.1$ such that stellar disc is stable against axisymmetric perturbation, but the self-gravity is still important \citep[for details, see][]{Toomre1964,JogSolomon1984}. 
\subsubsection{Disc stability against spiral density waves}

\begin{figure}
\centering
    \includegraphics[width=1.03\linewidth]{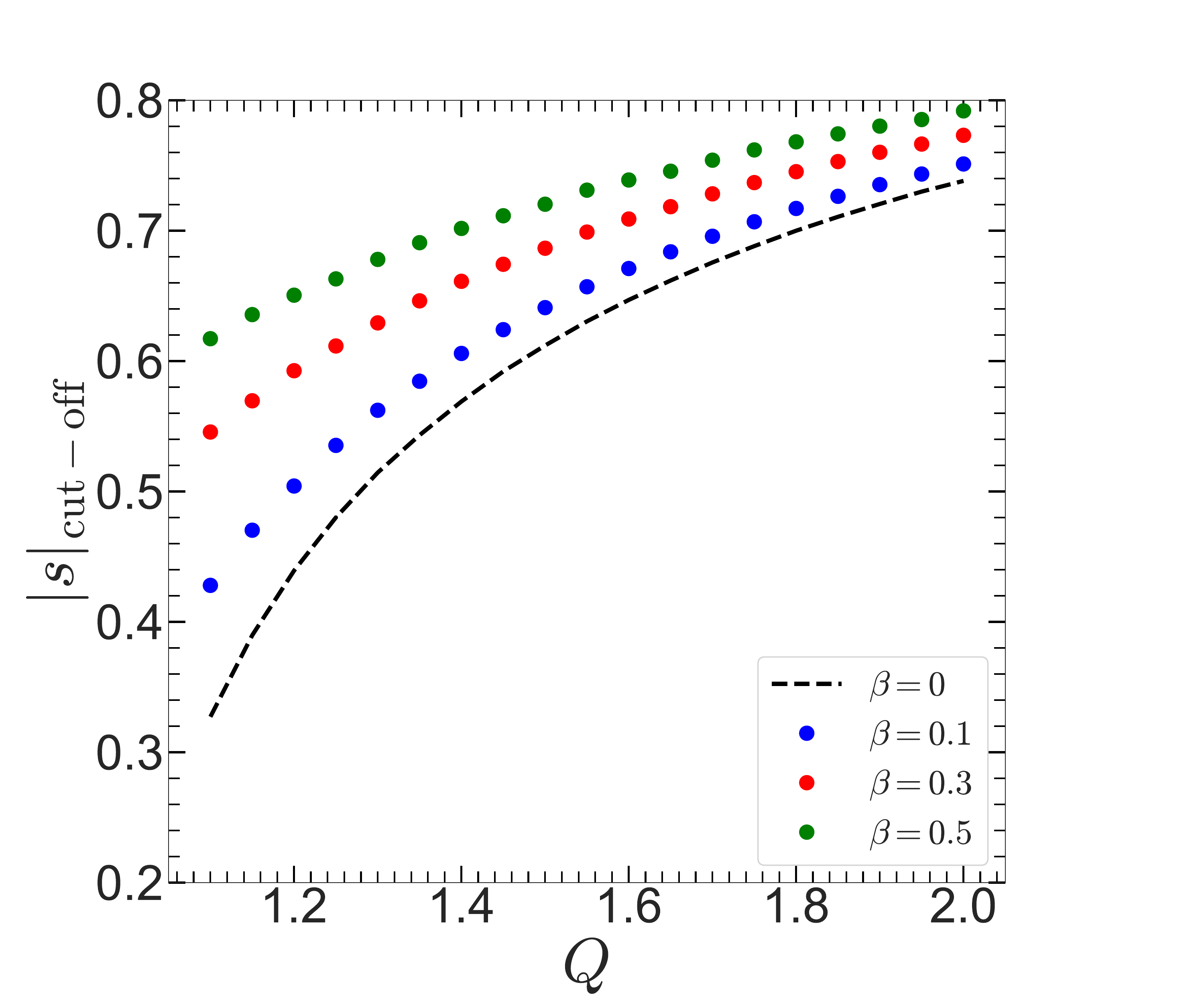}
\caption{\textbf{One-component stellar disc:} Variation of the $|s|_{\rm cut-off}$ values as a function of the Toomre $Q$ parameter are shown for different disc thickness ($\beta$). The black dashed line shows the corresponding $|s|_{\rm cut-off}$ values for a one-component stellar disc which is infinitesimally-thin. For a fixed Toomre $Q$, the  $|s|_{\rm cut-off}$ values increase monotonically with $\beta$.}
\label{fig:scutoff_onecomp}
\end{figure}

A visual inspection of Fig.~\ref{fig:dispreln_onecomp} reveals that with increasing thickness of the disc, the $|s|_{\rm cut-off}$ values increase steadily. This implies that the stellar disc becomes more and more stable against the non-axisymmetric perturbations. In order to study this more quantitatively, we calculate the $|s|_{\rm cut-off}$  values from the dispersion relation while systematically varying the Toomre $Q$ values from $1.1$ to $2$, and $\beta$ = 0.1, 0.3, and 0.5. The resulting variation of the  $|s|_{\rm cut-off}$ values for different $\beta$ and Toomre $Q$ parameter values are shown in Fig.~\ref{fig:scutoff_onecomp}. For reference, we also show the corresponding $|s|_{\rm cut-off}$ values for a one-component stellar disc with zero thickness (i. e., $\beta =0$).  As seen clearly from Fig.~\ref{fig:scutoff_onecomp}, for a fixed Toomre $Q$ parameter, the $|s|_{\rm cut-off}$ value increases monotonically with increasing $\beta$ values. Although this trend remains true for the whole range of Toomre $Q$ values considered here, the effect of increase in the $|s|_{\rm cut-off}$ value due to disc thickness is more prominent for lower Toomre $Q$ values considered here. For example, for $Q =1.1$, the $|s|_{\rm cut-off}$ values increase almost by a factor of $2$ on changing the $\beta$ from 0 (zero-thickness) to 0.5 ($\sim 385 $ pc); whereas the corresponding change is much smaller, by $\sim 7 \%$ for $Q=2$. This trend is not surprising, since for lower values of $Q$, the disc self-gravity is more relevant, and consequently the reduction in the self-gravity (due to finite-thickness) will be more pronounced as compared to the case with a higher Toomre $Q$ value. The physical implications of the increasing  $|s|_{\rm cut-off}$ value with thickness is discussed below. 
 \subsubsection{Effect of finite thickness on the allowed range of pattern speed values}
 
 As the  $|s|_{\rm cut-off}$ value increases with increasing thickness, this implies that the forbidden region  (i. e., the region between the CR and  $|s|_{\rm cut-off}$) also increases. In other words, introduction of a finite thickness of the disc helps to stabilise it against the non-axisymmetric perturbations.  The increase in the  $|s|_{\rm cut-off}$ value with increasing thickness has far-reaching impact on the pattern speed ($\Omega_{\rm p}$) value of a stable, spiral density wave. In a collisionless stellar disc, a spiral density wave exists only in those regions where 
 \begin{equation}
 \Omega -\kappa/2 \leq \Omega_{\rm p} \leq \Omega +\kappa/2\,,
 \label{eq:allowed_patternspeed}
 \end{equation}
\noindent is satisfied, and the equality holds only at the Lindblad resonance points \citep[for details see, e. g.,][]{BT08}. Also, at a certain radius $R$, and for a given pattern speed ($\Omega_{\rm p}$), the density wave will be \textit{stable} if the corresponding $|s| (= m|\Omega_{\rm p}-\Omega|/\kappa)$ is greater than the $|s|_{\rm cut-off}$ value, say $\alpha$, of the corresponding dispersion relation at that radius $R$ \citep[for detailed discussion see][]{BT08,GhoshJog2016}. In other words, for obtaining a \textit{stable} spiral density wave, with pattern speed $\Omega_{\rm p}$, the following condition needs to be satisfied
\begin{equation}
\Omega_{\rm p} \geq \Omega + \alpha \kappa/2 \hspace{0.2 cm} \mbox{or, } \hspace{0.2 cm} \Omega_{\rm p} \leq \Omega - \alpha \kappa/2\,,
\label{eq:allowed_stablepatternspeed}
\end{equation}
\noindent at radius $R$, depending on whether the radius $R$ falls outside the CR or inside the CR.
 Therefore, by combining Eqs. ~(\ref{eq:allowed_patternspeed}) and (\ref{eq:allowed_stablepatternspeed}), the allowed range of pattern speed ($\Omega_{\rm p}$) values for a stable, spiral density wave is
 \begin{equation}
 \Omega_{\rm p} \in [ \Omega -\kappa/2,  \Omega -\alpha \kappa/2] \hspace{0.2 cm} \mbox{or, } \hspace{0.2 cm} \Omega_{\rm p} \in [ \Omega +\alpha \kappa/2,  \Omega + \kappa/2]\,,
 \end{equation}
 \noindent depending on whether the radius $R$ is inside the corotation or outside the corotation. Now, as the value of $\alpha$ (or $|s|_{\rm cut-off}$) increases with thickness, the allowed range of pattern speed values for a stable spiral density wave become progressively narrower. This is another finding of this work.  Note that the inclusion of gas allows a larger range of allowed pattern speed values \citep{GhoshJog2016}. Thus, the effects of the gas and the disc thickness have an opposite effect on the range of  pattern speed values corresponding to a stationary (non-evanescent) spiral density wave. This is further discussed in Sect.~\ref{sec:results_twocomp}.
\par

\medskip

\subsubsection{Radial group transport and effect on longevity of spiral density waves}

Next, we study whether the radial group transport, and hence the longevity of a spiral density wave gets altered with the inclusion of finite thickness of the collisionless stellar disc. In classical density wave theory, the pattern speed ($\Omega_{\rm p}$) of the spiral arms is a free parameter \citep[e. g., see][]{LinandShu1964,LinandShu1966}. Observationally, the pattern speed of spiral density waves has been measured only for a few external galaxies \citep[e. g,][]{Fathietal2007,Fathietal2009} apart from the Milky Way. Therefore, driven by purely theoretical interest, we treat the pattern speed ($\Omega_{\rm p}$) as a free parameter. For investigating the effect of disc thickness on the group velocity, we first assume the pattern speed to be
$\Omega_{\rm p} = 12.5 \kms
$,
the same value as used in \citet{Toomre1969}. Also, we choose $Q =1.1$ here, and vary $\beta$ from 0 to 0.3. Now, using the definition of the dimensional quantity $s$ as given earlier in this section (i.e., $|s|= 2 (|\Omega_p - \Omega|)/\kappa $) and using $\Omega$ and $\kappa$ values for a particular galaxy at a given radius, and a given value of $\Omega_{\rm p}$ gives $|s|_{\rm obs}$, the observed value of $|s|$. This is shown by the horizontal line in Fig.~\ref{fig:dispreln_onecomp}. Then, we compute the group velocity ($v_g$) at $R = 8.27 \kpc$, from the slopes of the corresponding dispersion relations (as shown in Fig.~\ref{fig:dispreln_onecomp}) with varying $\beta$ values, the slope is obtained where the line $|s|_{\rm obs}$ intersects the dispersion relation.
The resulting values of the  group velocity ($v_g$) and the time that one such wavepacket (of density wave) would take to travel a distance of $10 \kpc$ are listed in Table~\ref{table:grpVel_onecomp}. Before proceeding to interpreting dynamical effect of thickness on the group velocity, the location where the slopes are being calculated, merits a discussion. For an assumed $\Omega_{\rm p}$ value and the values of $\Omega$, and $\kappa$ (set by the underlying mass distribution), the corresponding $|s|_{\rm obs}$ would intersect the dispersion relations at two points : one at the long wavelength branch (lower $x$ value), and another at a short wavelength branch (higher $x$ value). For a given $\Omega_{\rm p}$ value, we always calculate the group velocity in the short wavelength branch regime, as the WKB approximation works better there \citep[for details, see][]{BT08}.
\begin{table}[htbp]
\addtolength{\tabcolsep}{-1pt}
\centering
\caption{Group velocity for one-component stellar disc (with $Q =1.1$) for various thickness values, calculated at $R = 8.27 \kpc$.}
\begin{tabular}{cccccc}
\hline
$\Omega_{\rm p}$ & $|s|_{\rm obs}$ & $\beta$ & $ds/dx$ & $v_g$ & Time to travel\\
($\kmsk$) & & & & $(\kms)$ & $10 \kpc$ (Gyr)\\
\hline
10 & 0.92 & 0 & 0.032 & 1.3 & 7.5\\
&& 0.1 & 0.04 & 1.62 & 6.03\\
&& 0.2 & 0.045 & 1.83 & 5.3\\
&& 0.25 & 0.047 & 1.91 & 5.1\\
&& 0.3 & 0.05 & 2.03 & 4.8\\
\hline
12.5 & 0.8 & 0 & 0.107 & 4.34 & 2.25\\
&& 0.1 &  0.123 & 5 & 1.95\\
&& 0.2 & 0.137 & 5.56 & 1.75\\
&& 0.25 & 0.142 & 5.77 & 1.69\\
&& 0.3 & 0.143 & 5.81 & 1.68\\
\hline
15 & 0.68 & 0 & 0.181 & 7.35 & 1.32\\
&& 0.1 & 0.199 & 8.1 & 1.2\\
&& 0.2 & 0.203 & 8.25 & 1.18\\
&& 0.25 & 0.204 & 8.29 & 1.17\\
&& 0.3 & 0.205 & 8.3 & 1.16\\ 
\hline
\end{tabular}
\label{table:grpVel_onecomp}
\end{table}

From Table~\ref{table:grpVel_onecomp}, it is clearly seen that for $\Omega_{\rm p} =12.5 \kmsk$ and $Q=1.1$, the value of the group velocity increases steadily with increasing disk thickness ($\beta$). As the $\beta$ value changes from 0 to 0.3, the group velocity ($v_g$) increases by $\sim 35$ per cent. This implies that a wavepacket would take less time to reach the centre of the disc (and get absorbed eventually) for a disc with finite thickness when compared to that of an infinitesimally-thin disc. In other words, a spiral density wave would survive for a lesser time for a stellar disc with finite thickness when compared to an infinitesimally-thin stellar disc. Next, we choose $\Omega_{\rm p} =10 \kmsk$, and $15 \kmsk$, and recalculate the variation of the group velocity with changing disc thickness. Here also we choose $Q=1.1$. The corresponding results are also given in Table~\ref{table:grpVel_onecomp}. We find that, for these pattern speed values, the group velocity increases with the disc thickness, similar to the case of $\Omega_{\rm p} =12.5 \kmsk$. However, the amount by which the group velocity changes, varies with the assumed pattern speed values. To elaborate, the group velocity increases by $\sim 56$ percent for  $\Omega_{\rm p} =10 \kmsk$ whereas the the group velocity increases by $\sim 13$ percent for  $\Omega_{\rm p} =15 \kmsk$, once the thickness ($\beta$) is increased from 0 to 0.3. We also consider a higher Toomre $Q$ value, namely, $Q =1.5$ and study the variation of the group velocity with disc thickness. For the sake of brevity, the detailed variations in the group velocity values are not shown here. However, we find that, for an $Q=1.5$, and $\Omega_{\rm p} =12.5 \kmsk$, the corresponding group velocity ($v_g$) increases by $\sim 16$ per cent  for an increase in thickness from $\beta =0$ to $\beta =0.3$, as compared to $\sim 35$ per cent for $Q=1.1$ obtained earlier.

We note that the measured group velocity is critically dependent on the location where the slope is being measured along the dispersion relation. Nevertheless, an increment in the group velocity value with the increasing disc thickness is seen to be a generic phenomenon, as shown here. Lastly, we point out that for the one-component stellar disc case, we could not explore a higher Toomre $Q$ value and/or a higher pattern speed value, because for these cases, the measured $|s|_{\rm obs}$ value is found to be lower that the $|s|_{\rm cut-off}$ value. Consequently, the $|s|_{\rm obs}$ does not intersect the corresponding dispersion relation. In other words, for higher Toomre $Q$ values, and a higher pattern speed value, e.g., $\Omega_{\rm p} = 18 \kmsk$ does not admit a real solution in $k$ (or equivalently, $x$), and so we could not calculate the group velocity using Eq.~(\ref{eq:grp_velocity}). We will explore this parameter regime in the  next section where the interstellar gas is taken into account.

\subsubsection{Radial variation of the group transport}
\label{sec:RadVariation_grpTransport_onecomp}

So far, we have calculated the group velocity of a typical wavepacket, made of density wave, at a certain radius, say $R$, to study the effect of the finite thickness of a stellar disc on the longevity of the spiral density wave. 
We have so far considered the group transport at the solar neighbourhood $R = 8.27$ kpc. 
However, in reality, any spiral arm in a disc galaxy has a finite radial extent. Here, we study how the group velocity of a wavepacket changes at different radii for different thickness, and consequently how this affects the longevity of the spiral density wave.

We mention that, for an assumed flat rotation curve in the outer disc region, as done here, the values of $\kappa$ and $\Omega$ will change at different radial locations. Therefore, for an assumed value of $\Omega_{\rm p}$ which remains constant with respect to radius, the corresponding $|s|_{\rm obs}$ value would change at different radial location, so would the $x$ values where the $|s|_{\rm obs}$ cuts the local dispersion relation. To evaluate this, we first take $\Omega_{\rm p} = 12.5 \kmsk$ and $Q = 1.1$ (same as Table~\ref{table:grpVel_onecomp} which was done for $R=8.27$ kpc), and redo the group velocity calculation at $R = 6.27 \kpc$, $R = 7.27 \kpc$, and $R=9.27 \kpc$. The results are given in Table~\ref{table:grpVel_onecomp_radialvariation}.
 From Table~\ref{table:grpVel_onecomp_radialvariation}, it is clearly seen that at three different radial locations we considered here, the variation of the finite thickness from $\beta = 0$ to $\beta = 0.3$, leads to a monotonic increase of the group velocity of the density wavepacket. To express it more quantitatively, at $R = 6.27 \kpc$, the group velocity increases by $\sim 66.2$ percent when $\beta$ is varied from $0$ to $0.3$ whereas at $R = 7.27 \kpc$, and $R = 9.27 \kpc$, the corresponding group velocity increases by $\sim 49.8$ and $\sim 23.1$ percent, respectively (for the same $\beta$ variation). This trend is in concordance with that seen at $R = 8.27 \kpc$ (see Table~\ref{table:grpVel_onecomp}). In other words, the finite thickness has a similar (qualitative) effect on the group velocity as well as the longevity of spiral density waves. We checked this trend for the other considered pattern values, namely, $\Omega_{\rm p} =15 \kmsk$, and $\Omega_{\rm p} =10 \kmsk$ as well. We found a qualitative trend in the results similar to what is seen for $\Omega_{\rm p} = 12.5 \kmsk$, as long as the the dispersion relation admits a real solution in $x$ in the short wavelength brach for the corresponding $|s|_{\rm obs}$ value (for details, see previous section). For brevity they are not shown here. To conclude, the finite thickness of the stellar disc has a similar quenching effect on the longevity of spiral density waves at different radial locations (covering the radial extent of spirals). 
\begin{table}[htbp]
\addtolength{\tabcolsep}{-1.25pt}
\centering
\caption{Radial variation of group velocity for one-component stellar disc (with $Q =1.1, \Omega_{\rm p}= 12.5 \kmsk$), calculated at three radial locations.}
\begin{tabular}{cccccc}
\hline
$R$ & $|s|_{\rm obs}$ & $\beta$ & $ds/dx$ & $v_g$ & Time to travel\\
($\kpc$) & & & & $(\kms)$ & $10 \kpc$ (Gyr)\\
\hline
6.27 & 0.94 & 0 & 0.021 & 1.14 & 8.5\\
&& 0.1 & 0.027 & 1.46 & 6.7\\
&& 0.2 & 0.032 & 1.71 & 5.7\\
&& 0.25 & 0.034 & 1.82 & 5.35\\
&& 0.3 & 0.035 & 1.9 & 5.1\\
\hline
7.27 & 0.87 & 0 & 0.06 & 2.8 & 3.5\\
&& 0.1 &  0.074 & 3.42 & 2.85\\
&& 0.2 & 0.083 & 3.8 & 2.56\\
&& 0.25 & 0.086 & 3.9 & 2.45\\
&& 0.3 & 0.09 & 4.2 & 2.32\\
\hline
9.27 & 0.72 & 0 & 0.156 & 5.6 & 1.73\\
&& 0.1 & 0.178 & 6.4 & 1.51\\
&& 0.2 & 0.184 & 6.7 & 1.45\\
&& 0.25 & 0.188 & 6.8 & 1.43\\
&& 0.3 & 0.192 & 6.9 & 1.41\\ 
\hline
\end{tabular}
\label{table:grpVel_onecomp_radialvariation}
\end{table}

\subsection{Effect of finite height on two-component star-gas system}
\label{sec:results_twocomp}
In the earlier section, we show that the inclusion of finite thickness of the stellar disc makes it more stable against the non-axisymmetric perturbations, and increases the group velocity of a wavepacket, thereby decreasing the longevity of the spiral density wave. Further, \citet{GhoshJog2015} showed that for a gravitationally-coupled star-gas system where both the stellar and the gas discs are \textit{infinitesimally-thin}, the inclusion of gas helps the spiral density waves to sustain for a longer time. 
Therefore, it is natural to investigate how the longevity of the spiral density wave is affected in a stars plus gas system with finite disc thickness.
\begin{figure}
\centering
    \includegraphics[width=1.0\linewidth]{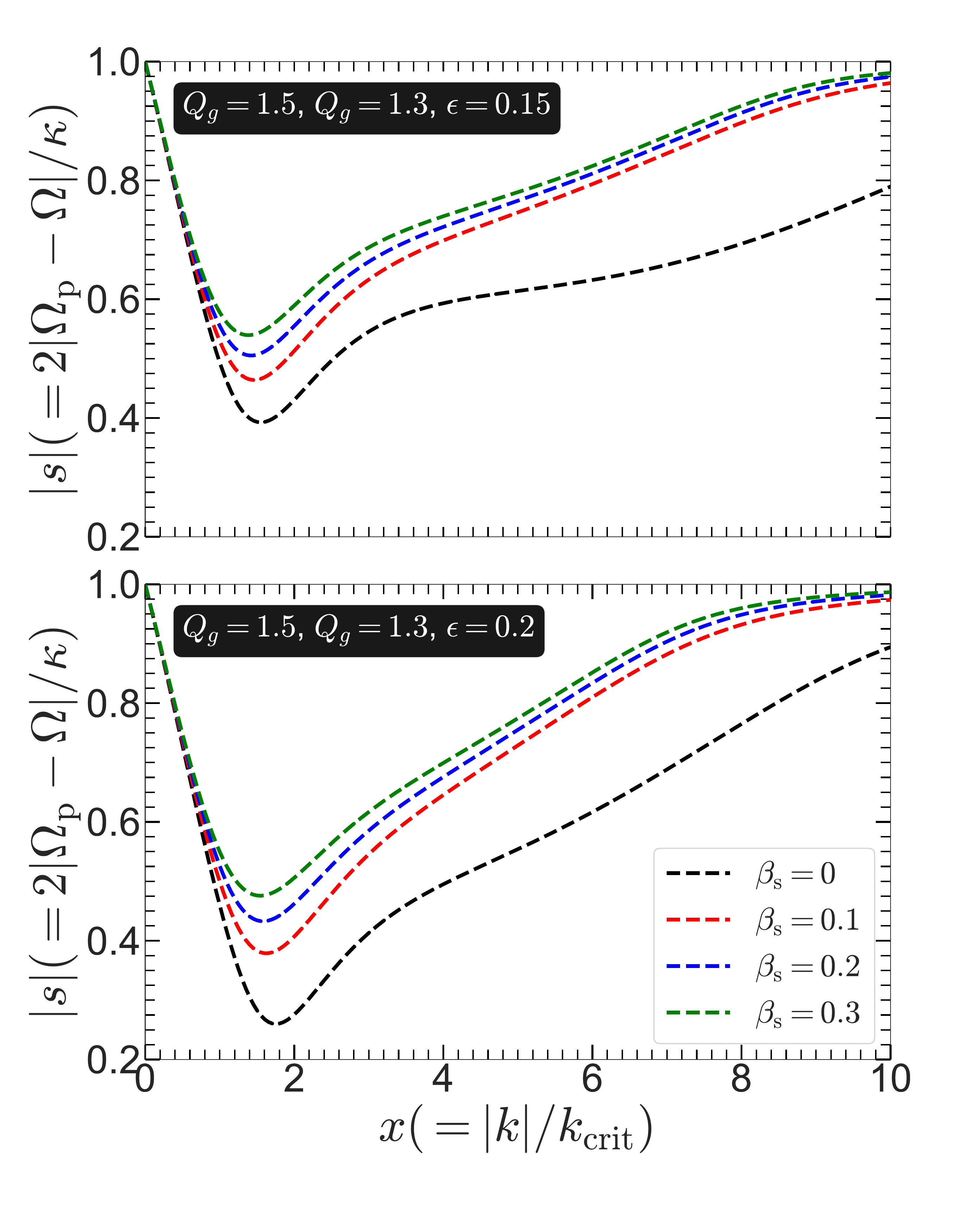}
\caption{Dispersion relations for the two-component stars plus gas system with finite thickness (Eq.~(\ref{disp-twocom})) are shown for $Q_{\rm s} =1.5$, $Q_{\rm g} =1.3$, and for different gas-fractions : $\epsilon = 0.15$ (top panel), $\epsilon = 0.2$ (bottom panel). The thickness of the stellar disc ($\beta_{\rm s}$) is varied from 0.1 to 0.3, whereas the thickness of the gas disc ($\beta_{\rm g}$) is kept fixed at 0.1 throughout all cases shown here. For comparison, the corresponding dispersion relation for the zero-thickness stars plus gas system is also shown (black dashed line).}
\label{fig:dispreln_twocomp_example}
\end{figure}
\begin{figure*}
\centering
    \includegraphics[width=1.02\linewidth]{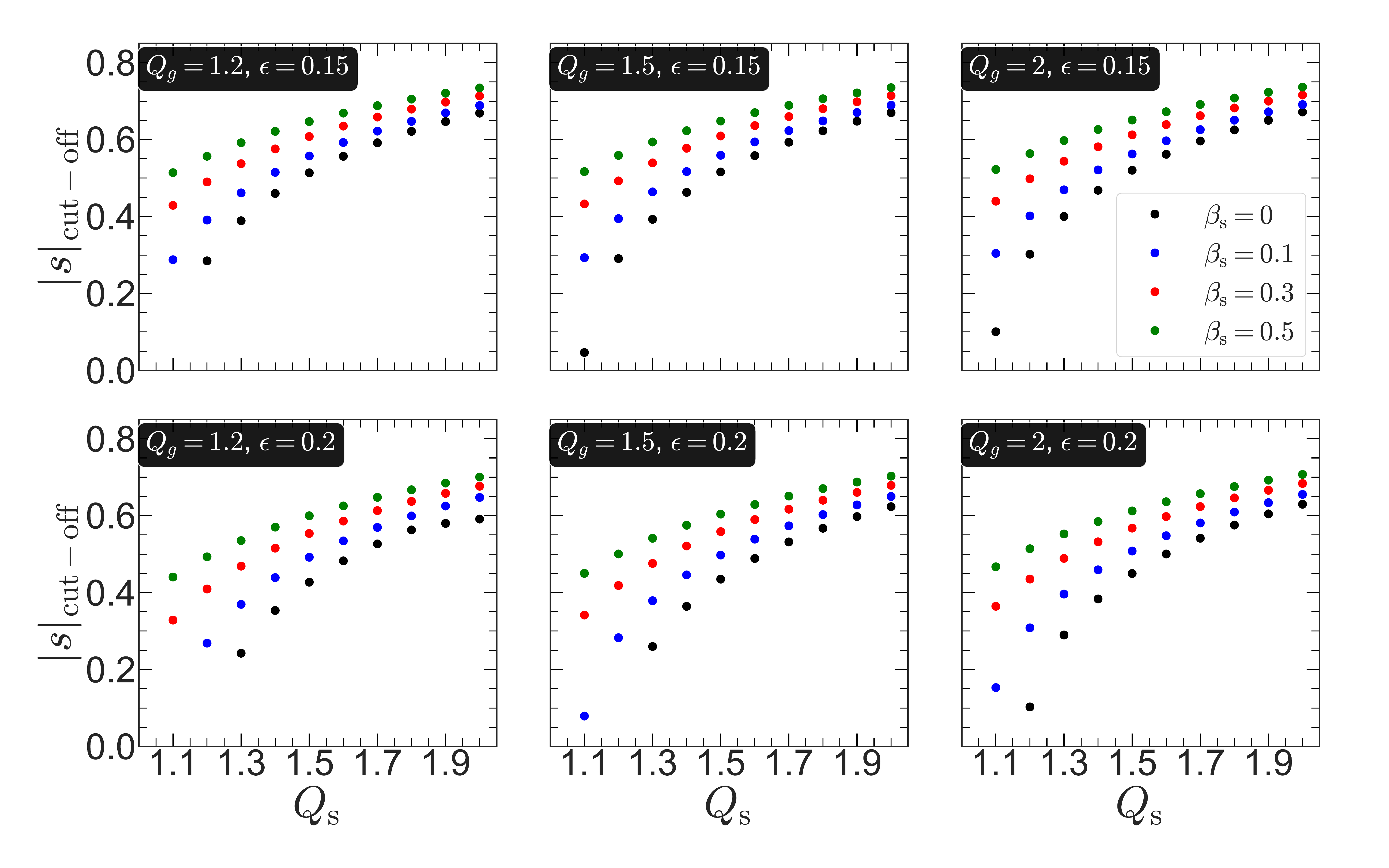}
\caption{\textbf{Two-component stars plus gas system :} Variation of the $|s|_{\rm cut-off}$ values as a function of Toomre $Q$ parameter for the stellar disc ($Q_{\rm s}$) are shown for different stellar disc thickness ($\beta_{\rm s}$), and for various Toomre $Q$ parameter for the gas disc ($Q_{\rm g}$), and gas-fraction ($\epsilon$). The thickness of the gas disc ($\beta_{\rm g}$) is kept fixed at 0.1 throughout all 
the cases shown here.}
\label{fig:scutoff_twocomp}
\end{figure*}
\par
To achieve that, first we calculate the dispersion relation (Eq.~(\ref{disp-twocom})) for the gravitationally-coupled star-gas system for different thickness of stellar disc ($\beta_{\rm s}$) while assuming $Q_{\rm s} = 1.5$, $Q_{\rm g} = 1.3$, and taking two values of $\epsilon$, namely, $0.15$, and $0.2$. We mention that, the thickness of the gas disc ($\beta_{\rm g}$) is kept fixed at 0.1 throughout this paper, unless stated otherwise. The resulting dispersion relations are shown in Fig.~\ref{fig:dispreln_twocomp_example}. A visual inspection reveals two broad trends, namely, for a fixed set of three values $(Q_{\rm s}, Q_{\rm g}, \epsilon)$, the value of $|s|_{\rm cut-off}$ increases monotonically with the increase of thickness of the stellar disc. This trend is consistent with the findings for the one-component stellar disc. However, we notice that, with increasing gas fraction ($\epsilon$), the increment in the $|s|_{\rm cut-off}$ value with thickness, is less (compare top and bottom panels of Fig.~\ref{fig:dispreln_twocomp_example}). Secondly, the dispersion relation in the short-wavelength branch become increasingly flat with the increase of disc thickness, when compared with the same for the infinitesimally-thin star-gas system. 
This holds true for both the gas-fraction values considered here. 
\subsubsection{Disc stability against spiral waves, and allowed pattern speed values}

In order to probe the joint effect of the disc thickness and the inclusion of gas on the variation of $|s|_{\rm cut-off}$ values, we systematically calculate the relevant dispersion relations (using Eq.~(\ref{disp-twocom})) for a wide range of $Q_{\rm s}$, $Q_{\rm g}$, $\beta_{\rm s}$, and $\epsilon$. Then we compute the corresponding $|s|_{\rm cut-off}$ values. The resulting variation of $|s|_{\rm cut-off}$ values are shown in Fig.~\ref{fig:scutoff_twocomp}. Fig.~\ref{fig:scutoff_twocomp} clearly demonstrates the generic trend that, for a fixed value of $(Q_{\rm s}, Q_{\rm g}, \epsilon)$, the $|s|_{\rm cut-off}$ increases monotonically while the disc thickness ($\beta_{s}$) is varied from 0 to 0.5. However, the variation in the $|s|_{\rm cut-off}$ values is more for a lower $Q_{\rm s}$ value (say 1.2) when compared with a higher $Q_{\rm s}$ value (say, $\sim 2$). These findings are in compliance with the earlier finding for the one-component stellar disc. Next, we study how the variation  in the $|s|_{\rm cut-off}$ values with thickness, gets affected by the inclusion of the interstellar gas. We find that, for a fixed value of ($Q_{\rm g}$, $\beta_{\rm s}$), the $|s|_{\rm cut-off}$ value for a higher gas-fraction ($\epsilon$) is always lower when compared with the same for a lower  gas-fraction (compare top and bottom panels of Fig.~\ref{fig:scutoff_twocomp}). This trend holds true for the whole range of $Q_{\rm s}$ and $Q_{\rm g}$values considered here, although the trend is more prominent for the smaller values of $Q_{\rm s}$. The physical reason behind this trend is when more gas present in the system (a lower $Q_{\rm g}$ value and a higher $\epsilon$ value), the destabilising effect of the interstellar gas wins over the stabilising effect of the finite thickness. This is particularly true when the $Q_{\rm s}$ and $Q_{\rm g}$ values are smaller (say close to 1) as the self-gravity of the stars-gas system would be important to determine the dynamical state of the system. Thus, the net stability of the gravitationally-coupled stars plus gas system is determined by the joint (mutually opposite) effects of the interstellar gas and the finite disc thickness. Nevertheless, the destabilising effect of the interstellar gas gets reduced (at least, partially) by the stabilising effect of the finite disc thickness in all cases shown here.Thus, the net range of allowed pattern speed values is higher for a two-component, finite height disc than a stars-alone, thin disk; but less than the range allowed for a two-component, thin disc.
\par
\medskip

\subsubsection{Radial Group transport and effect on longevity of spiral waves}

Finally, we turn to probe the joint effects of the disc finite thickness and the interstellar gas on the group velocity of the spiral density wave in a gravitationally-coupled stars-gas system. For the sake of theoretical interest, we choose a smaller value of $Q_{\rm s}$, say 1.3. Also, for the moment, we treat $Q_{\rm s}$ and $\epsilon$ as free parameters, and $Q_{\rm g}$ is set by these two values via the relation $Q_{\rm g}$ = $(0.306 Q_{\rm s}) (1-\epsilon)/\epsilon$ \citep[for details see][]{GhoshJog2015}. Later, we will relax this constraint and will treat all three parameters, namely, $Q_{\rm s}$, $Q_{\rm g}$, and $\epsilon$ as free parameters while studying the group velocity.

To study how the group velocity changes, first we assume a $\Omega_{\rm p} = 12.5 \kmsk$ which in turn, gives $|s|_{\rm obs} = 0.8$ at $R = 8.27 \kpc$. Then we compute the group velocity from the slopes of the dispersion relations (Eq.~(\ref{disp-twocom})) for $Q_{\rm s} =1.3$ and $\beta_{\rm g} =0.1$ while varying the gas-fraction ($\epsilon$) from 0.1 to 0.25, and disc thickness ($\beta_{\rm s}$) from 0 to 0.3. The resulting variations of the slopes and the group velocity values are listed in Table~\ref{table:grpVel_twocomp}. As evident from Table~\ref{table:grpVel_twocomp}, the value of the group velocity increases with the increase of disc thickness, and this trend holds true for all values of gas-fraction considered here. However, there is subtle change with the increasing gas-fraction. To elaborate, when the gas-fraction is assumed to be 0.1, the group velocity increases by $\sim 35$ percent for a variation of $0-0.3$ in the $\beta$ values. However, for a higher gas-fraction value (say $\epsilon = 0.2$),  the group velocity increases only by $\sim 12$ percent for the same variation of $\beta$ values from 0 to 0.3. Further, when we choose an even higher value of gas-fraction (say, $\epsilon = 0.25$), we find a negligible increase in the group velocity ($\sim 2$ percent) when we vary $\beta$ from 0 to 0.3. The physical reason is as we choose progressively lower values of $Q_{\rm g}$ and higher values of $\epsilon$, the overwhelming effect of the interstellar gas in decreasing the group velocity wins over the opposite effect of disc thickness on the group velocity values. In other words, when the self-gravity, and the low dispersion of interstellar gas dominates over the reduction in the self-gravity due to the finite thickness, the group velocity of a wavepacket, made of spiral density waves, is predominantly set by the effect of the interstellar gas.  Next, we choose a higher pattern speed value, namely, $\Omega_{\rm p} = 15 \kmsk$ which in turn, produces $|s|_{\rm obs} = 0.68$  $R = 8.27 \kpc$. Then, we study how the group velocity changes as we vary simultaneously the gas-fraction, and the disc finite thickness. We find that, when $\epsilon =0.1$, the group velocity increases by $\sim 20$ percent when the disc thickness is increased from 0 to 0.3. However, when the gas-fraction is changed to a higher value (say $\epsilon =0.2$), the group velocity increases only by $\sim 14$ percent when $\beta_{\rm s}$ is varied from 0 to 0.3. This trend is similar to what we found for the $\Omega_{\rm p} = 12.5 \kmsk$. Thus, the change in group velocity in presence of the disc thickness and the interstellar gas is a complex process, as these two physical factors have opposite effect on the group velocity. The net change in the group velocity is set by the relative dominance of the effects of the interstellar gas and the disc thickness.
\begin{table}[htbp]
\addtolength{\tabcolsep}{-1.75pt}
\centering
\caption{Group velocity for two-component star-gas system (with $Q_{\rm s} =1.3$) for various thickness values, calculated at $R = 8.27 \kpc$.}
\begin{tabular}{ccccccc}

\hline
$\Omega_{\rm p}$ & $|s|_{\rm obs}$ & $\epsilon$ & $\beta$ & $ds/dx$ & $v_g$ & Time to \\
($\kms$ & & & & &  $(\kms)$ &  travel $10 \kpc$ \\
 kpc$^{-1}$) &&&&&&  (Gyr)\\
\hline
12.5 & 0.8 & 0.1 & 0 & 0.081 & 3.3 & 2.95\\
&&& 0.1 & 0.1 & 4.1 & 2.37\\
&&& 0.2 & 0.104 & 4.2 & 2.32\\
&&& 0.3 & 0.11 & 4.5 & 2.16\\
\hline
&& 0.15 & 0 & 0.067 & 2.72 & 3.58\\
&&& 0.1 & 0.082 & 3.35 & 2.9\\
&&& 0.2 & 0.083 & 3.4 & 2.86\\
&&& 0.3 & 0.084 & 3.41 & 2.85\\
\hline
&& 0.2 & 0 & 0.074 & 3 & 3.24\\
&&& 0.1 & 0.083 & 3.37 & 2.89\\
&&& 0.2 & 0.0832 & 3.38 & 2.88\\
&&& 0.3 & 0.0833 & 3.39 & 2.87\\
\hline
&& 0.25 & 0 & - & - & -\\
&&& 0.1 & 0.093 & 3.4 & 2.86\\
&&& 0.2 & 0.095 & 3.42 & 2.84\\
&&& 0.3 & 0.095 & 3.42 & 2.84\\
\hline
\end{tabular}
{* The Toomre $Q$ for gas disc ($Q_{\rm g}$) is set by the values of $Q_{\rm s}$ and $\epsilon$, for details see text. The thickness of the gas disc $\beta_{\rm g}$ is fixed at 0.1 for all cases.}
\label{table:grpVel_twocomp}
\end{table}
\par
To study further the net effect of the interstellar gas and the disc thickness where the parameters $Q_{\rm s}$, $Q_{\rm g}$, and $\epsilon$ are treated as free parameters (unlike the previous case), we choose a case where $Q_{\rm s} = 1.3$, and $Q_{\rm g} = 1.3$. For these assumed parameters, and for $\epsilon =0.2$ and $\Omega_{\rm p} = 15 \kmsk$, the group velocity increases by $\sim 25$ percent when $\beta_{\rm s}$ is varied from 0 to 0.3. However, for a higher $\epsilon = 0.25$, the change in the group velocity is negligible when $\beta_{\rm s}$ is varied from 0 to 0.3. This trend further demonstrate the mutual interplay of the (opposite) effects induced by the interstellar gas and the disc finite thickness. Lastly, we assume an even higher value of the pattern speed, namely, $\Omega_{\rm p} = 18 \kmsk$ which in turn, yields an $|s|_{\rm obs} = 0.53$. For the current assumed parameter space ($Q_{\rm s}$, $Q_{\rm g}$, $\epsilon$, $\beta_{\rm s}$), it is possible to obtain a real solution in $k$ (or alternatively, in $x$) for the $\Omega_{\rm p} = 18 \kmsk$; thereby allowing us to probe the mutual effect of the gas and the thickness for a realistic value (or close to observationally reported values) of the $\Omega_{\rm p}$. Here also, we find the same trend in the change of the group velocity value as we increase the value of $\beta_{\rm s}$. More quantitatively, when $\epsilon =0.2$, the group velocity increases by $\sim 11$ per cent as we vary $\beta_{\rm s}$ from 0 to 0.3. However, the change in the group velocity is found to be negligible when $\epsilon$ is set to $0.25$ while $\beta_{\rm s}$ is varied by a same amount as previous case. This further accentuates the complex mutual interplay of the opposite effects of the disc thickness and the interstellar gas on the resulting group velocity values.

Lastly, we explore the mutual effect of the disc thickness and the interstellar gas for some higher values of $Q_{\rm s}$ while varying the thickness of the stellar disc up to $\sim 400$ pc (equivalently, $\beta_{\rm s} \sim 0.5$). This is close to the observed values of the disc thickness in the external galaxies \citep[e. g., see][]{degrijs1997}. For that, we first choose $Q_{\rm s} =1.6$, $Q_{\rm g} = 1.5$, and $\epsilon = 0.2$, and then vary the disc thickness ($\beta_{\rm s}$) from 0 to 0.5. We find that, for this chosen set of values for ($Q_{\rm s}$, $Q_{\rm g}$, $\epsilon$), and for an assumed $\Omega_{\rm p}$ of $15 \kmsk$, the group velocity increases monotonically with increasing disc thickness. We estimate that, for a variation of the disc thickness from 0 to 0.5, the group velocity increases by $\sim 37$ per cent. Further, we choose another set of values for ($Q_{\rm s}$, $Q_{\rm g}$, $\epsilon$), namely, $Q_{\rm s} =2$, $Q_{\rm g} =2$, and $\epsilon = 0.2$. This set of chosen values is typical for the outer regions of the discs of Magellanic-type irregular galaxies where the gas fractions are high \citep[e. g., see][]{GallagherandHunter1984,Jog1992}. We then choose a $\Omega_{\rm p} = 12.5 \kmsk$, and vary the  disc thickness from 0 to 0.5. We find that, for this case, the group velocity increases by $\sim 13 $ percent as the disc thickness is varied from 0 to 0.5. Note that, for this case, we could not use a higher pattern speed value, say $\Omega_{\rm p} = 15 \kmsk$ or higher as the group velocity approach cannot be applied here. The reason is, the $|s|_{\rm obs}$ values corresponding to these higher pattern speed values are almost always less than the $|s|_{\rm cut-off}$ values obtained from the dispersion relations corresponding to $Q_{\rm s} =2$, $Q_{\rm g} =2$, and $\epsilon = 0.2$, and $\beta_{\rm s} = 0.2-0.5$. 

\subsubsection{Coverage on the parameter space}

Thus, to conclude, we study the  joint effect of the disc thickness and the interstellar gas on the stability of the disc against non-axisymmetric perturbations as well as on the longevity of the $m=2$ spiral density waves. By exploring a range of parameter space, we demonstrate the complex nature of the mutually opposite effect of the disc thickness and the interstellar gas. We mention that, while studying this mutually opposite effect on the longevity of the spiral density wave, we could not carry out a systematic search in the parameter space, 
unlike the case of disc stability (or equivalently, the variation of $|s|_{\rm cut-off}$).  The reasons are as follows:
first, the group velocity is obtained using the slope of the local dispersion relation graphically \citep[as in][]{Toomre1969}. Although this approach allows us to calculate the group velocity conveniently, this is a 
non-robust procedure \citep[for details, see][]{Toomre1969,GhoshJog2015}. The slope depends critically on the exact location of $x$ where the $|s|_{\rm obs}$ (corresponding to a $\Omega_{\rm p}$ value) intersects the dispersion relation. In other words, it is a local value. Secondly, for a two-component system, the dispersion relation is a fourth-order polynomial \citep[e.g.][]{JogSolomon1984}. Thus for certain choice of parameters, if the  $|s|_{\rm obs}$ intersects the dispersion relation in the region of high $x$ that falls in the region of second minimum, the slope would tend to flatten regardless of the fact whether or not the solution in $x$, corresponding to that $|s|_{\rm obs}$ value, is real or imaginary. Hence, while normally a higher $\beta$ or thickness leads to a steeper slope, in such cases the flattening effect due to the $k^4$ behaviour may dominate and hence one may get a smaller slope for higher $\beta$, and hence leading to a smaller value of the group velocity. This is opposite to the typical dependence on disc thickness as shown above where a higher $\beta$ was shown to result in a higher group velocity. In such cases, the effect of disc thickness would not oppose, instead it would show a similar trend to the effect of gas. This complex, mixed behaviour is more likely to be seen at high gas fraction \citep[see][]{JogSolomon1984} or high $|s|_{\rm obs}$. This could contribute to a diverse and complex dynamical behaviour, and hence the results obtained from varying the $\beta$ parameter for a fixed set of ($Q_{\rm s}$, $Q_{\rm g}$, $\epsilon$) values have to be interpreted with caution. 
\par
Also, so far we kept the thickness of the gas disc, $\beta_{\rm g} = 0.1$ while calculating the dispersion relations for a two-component stars-plus-gas system. Using the $k_{\rm crit}$ value mentioned in section~\ref{sec:One-component fluid disc with finite thickness}, this reduces to a $h_{\rm g} \sim 70$ pc which is typical scale-height for the molecular hydrogen gas \citep[e.g., ][]{ScovilleandSanders1987}. On the other hand, neutral hydrogen (H~{\sc i}) shows a typical value for the scale-height as $\sim 150$ pc \citep[e.g., ][]{Lockman1984} which corresponds to $\beta_{\rm g} = 0.2$. Here, we briefly state what happens to the findings mentioned above when we set $\beta_{\rm g} = 0.2$ instead of $\beta_{\rm g} = 0.1$. We choose $Q_{\rm s} = 1.3$, $Q_{\rm g} = 1.3$, and $\epsilon = 0.2$, a case already explored in section~\ref{sec:results_twocomp}. We then set $\beta_{\rm g} = 0.2$, and vary $\beta_{\rm s}$ from 0-0.3, as before, to see the change in the radial group velocity corresponding to $\Omega_{\rm p} = 15 \kmsk$, and $\Omega_{\rm p} = 18 \kmsk$. For a fixed $\beta_{\rm s}$ and $\Omega_{\rm p}$ values, the group velocity is seen to increase by $\sim 13-55$ per cent for $\beta_{\rm g} = 0.2$ when compared with that for $\beta_{\rm g} = 0.1$. This shows that finite thickness of the gas disc also increases the group velocity for the joint disc (and consequently decreases the longevity) in a similar fashion as the thickness of the stellar disc. We also considered a few other set of ($Q_{\rm s}$, $Q_{\rm g}$, $\epsilon$) values to test this. We find the general trend of an increasing group velocity with an increase in thickness of the gas disc, for brevity we have not shown these cases here.
\par
Nevertheless, we stress that, in general, the inclusion of thickness of the stellar disc and the interstellar gas have opposing effects on the disc stability against non-axisymmetric perturbations and the radial group transport, as illustrated in terms of the typical examples in this section (also see Table~\ref{table:grpVel_twocomp}). In general, for the observed disc thickness and gas fraction values, as we have discussed in this section, the quenching effect of the height does not overcome (completely) the supporting role played by the  gas, and hence a disc would still be expected to host non-axisymmetric features whose longevity will be supported by the gas. 
 
\subsubsection{Radial variation of the group transport for a two-component disc}
\label{sec:RadVariation_grpTransport_twocomp}

As before, we study here the radial variation of the effect of finite thickness on the group velocity of a wavepacket, made of density wave, for a gravitationally-coupled stars-gas system. As before, we choose three radial locations at $R = 6.27 \kpc$,  $R = 7.27 \kpc$, and $R = 9.27 \kpc$. Then we assume a $\Omega_{\rm p} = 12.5 \kmsk$, and recalculate the group velocities for different disc thickness and the gas fraction values while choosing $Q_{\rm s}$ and $Q_{\rm g}$ values identical to what used for Table~\ref{table:grpVel_twocomp}. Here also, we find a similar qualitative trend of mutually-opposing effects of the finite thickness and the interstellar gas on the group velocity as well as the longevity of the spiral density waves. To elaborate, at $R = 6.27 \kpc$, and for $\epsilon =0.1$, and $\Omega_{\rm p} = 12.5 \kmsk$, the group velocity increases by $\sim 63.6$ percent when $\beta_{\rm s}$ is varied from $0$ to $0.3$. As the gas fraction ($\epsilon$) increases, the increment in the group velocity due to a variation of $\beta$ from 0 to 0.3, starts to diminish monotonically. For $\epsilon = 0.2$, the corresponding increment in the group velocity becomes $\sim 37.8$ percent, and for $\epsilon = 0.25$, the increment in the group velocity becomes $\sim 33.7$ percent (for the same variation in $\beta$ from 0 to 0.3). In other words, when the gas fraction becomes higher, the dynamical effect of the interstellar gas in supporting the spiral density wave starts to dominate over the quenching effect of the finite thickness. This trend is similar to what was shown for $R = 8.27 \kpc$ (see Table~\ref{table:grpVel_twocomp}). We further checked for other $\Omega_{\rm p}$ values as well as for other two radial locations considered here. We found that as long as the dispersion relation admits a real solution for $x$ in the long wavelength branch for the corresponding $|s|_{\rm obs}$ value, and the $k^4$ behaviour does not interfere in the locations where the group velocity is being calculated (for details, see the discussion in previous section), we obtained a similar qualitative effect of the finite thickness on the group velocity and on the longevity of the spiral density wave. For brevity, they are not shown here. Therefore, the finite thickness of the stellar disc has a similar quenching effect on the longevity of spiral density waves at different radial locations (covering the radial extent of spirals) for the gravitationally-coupled two-component disc as well.

\section{Discussion}
\label{sec:discussion}
Here, we discuss a couple of points relevant for this work.

\begin{itemize}

\item{ At large values of scale height, its effect begins to dominate the effect of gas, hence the net effect is to increase the group velocity which would lead to a shorter lifetime of the density wave.  A typical galaxy is known to show a flaring stellar disc as shown observationally by \citet{degrijs1997}. 
This feature occurs naturally as shown by the theoretical modelling of a multi-component galactic disc \citep{NarayanJog2002a}.  The Milky Way stellar disc also shows flaring as seen observationally from several surveys, e.g., 2MASS \citep{Momanyetal2006}, SEGUE \citep{Lopezetal2014}, LAMOST \citep{Wangetal2018} as well as LAMOST plus Gaia \citep{Yuetal2021}. Such flaring was explained by theoretical modelling of a multi-component disc \citep{NarayanJog2002}, especially for the outer Galaxy by \citet{SarkarandJog2018}. This could explain in a generic way why spiral features are rare in outer regions even though such regions tend to be gas-rich.}

\item{The formulation presented here employs the linear perturbation theory. However, the growth of the spirals in a real galaxy can enter the non-linear regime as well. The study by \citet{Donghia2013} demonstrated that the non-linear response of the disc to the non-axisymmetric perturbations can alter the life-time of spiral instability in a disc galaxy. While the study presented here, clearly demonstrates the mutually-opposite effect of the interstellar gas and the disc thickness on the disc stability and persistence of spiral density wave, it will be worth pursuing this effect in numerical simulations of disc galaxies. Also, we mention that, here the spiral structure is modelled as a two-dimensional structure for simplicity, as is typically done.
However, some of the recent self-consistent simulations of growth of the spirals in disc galaxies \citep[e. g., see][]{Debattista2014,Ghoshetal2020} show that the spirals can well be vertically extended.}

\end{itemize}
\section{Conclusion}
\label{sec:conclusion}

In summary, we investigated the dynamical impact of the disc thickness on the disc stability, and the longevity of the spiral density wave. For that, we derived the dispersion relations in the WKB approximation for a collisionless stellar disc with finite thickness as well as for a gravitationally coupled stars-plus-gas system with different thickness for the stars and the gas discs. The main findings are mentioned below.

\begin{itemize}

\item{The inclusion of finite thickness effectively reduces the self-gravity of the disc which in turn makes the disc more stable against the non-axisymmetric perturbations. This result in case of a one-component stellar disc, holds true for the whole range of Toomre-$Q$ parameter ($1.2-2$) and the disc thickness ($\sim 50 - 400$ pc) considered in this work. This stabilising effect is more prominent for lower values of Toomre-$Q$ where the self-gravity of the disc is  more important.}

\item{The stabilising effect of the finite disc thickness has consequences in setting up the allowed range of pattern speed values for which the system allows a non-evanescent spiral density wave. With increasing disc thickness, this allowed range of pattern speed values gets progressively narrower, especially for lower values of Toomre-$Q$ parameter. For a joint stars-plus-gas system, the effect of gas is to increase the range of pattern speed values while the thickness has an opposite effect. Typically, the net range of allowed pattern speed values is larger for a two-component, finite height disc than a stars-alone, infinitesimally-thin disc.}

\item{For a one-component stellar disc, the group velocity of a wavepacket increases monotonically with the increment of disc thickness, thereby implying a progressively shorter life-time for the spiral density wave. For the same change in disc thickness by $\sim 250$ pc, the reduction in the life-time can vary from $\sim 10-60$ per cent when compared to the case of an infinitesimally-thin stellar disc; and this reduction depends on the assumed pattern speed values and the Toomre-$Q$ parameter. In a two-component system, as the gas fraction is increased, the increase in the group velocity caused by the effect of finite height is reduced. Hence the inclusion of gas opposes the reduction in lifetime arising due to the finite thickness.}

\item{Even in the presence of the interstellar gas, the disc thickness tends to stabilise the two-component (stars-gas) system against the non-axisymmetric perturbation, and this holds true for the whole parameter space ($Q_{\rm s} = 1.2 -2$, $Q_{\rm g} = 1.2 -2$, $\epsilon \sim 0.10-0.25$) considered here. However, when the dynamical effect of gas is important (e.g., lower values of $Q_{\rm g}$, and/or higher $\epsilon$ values), the stabilising effect of the disc thickness becomes minimal.}

\item{The disc thickness tends to diminish the life-time of the spiral density waves, even when the gas is present. For the same change in disc thickness by $\sim 400$ pc, the amount of reduction in the life-time can vary from $\sim 5- 40$ per cent when compared to an infinitesimally-thin stars-plus-gas system, depending on the assumed pattern speed and the $Q_{\rm s}$, $Q_{\rm g}$ and gas-fraction ($\epsilon$) values. However, for gas-rich systems ($\epsilon \geq 0.25$), the effect of finite disc thickness is shown to be negligible. }

\end{itemize}

Thus, to conclude, we demonstrate in this paper that the interstellar gas and the disc thickness have an opposite dynamical effect on the disc instability and the persistence of spiral density waves. While the gas makes a galactic disc more susceptible against non-axisymmetric perturbations and helps the spiral density wave to survive for a longer time-scale, the disc thickness influences the system in an opposite manner. Consequently, the net dynamical effect is set by the relative dominance of these two physical factors. For the broad range of parameter space we considered, the quenching effect of the height does not completely suppress the supporting role of the  gas, and hence a disc would still be expected to host non-axisymmetric features whose longevity will be supported by the gas.

\section*{Acknowledgements}
We thank the anonymous referee for useful comments which helped to improve this paper. C. J.  thanks the DST, Government of India for support via a J.C. Bose fellowship (SB/S2/JCB-31/2014).

\bibliographystyle{aa} 
\bibliography{my_ref} 

\begin{appendix}

\section{Behaviour of the Reduction factor}
\label{appen:reduc}

\begin{figure}
\centering
    \includegraphics[width=1.05\linewidth]{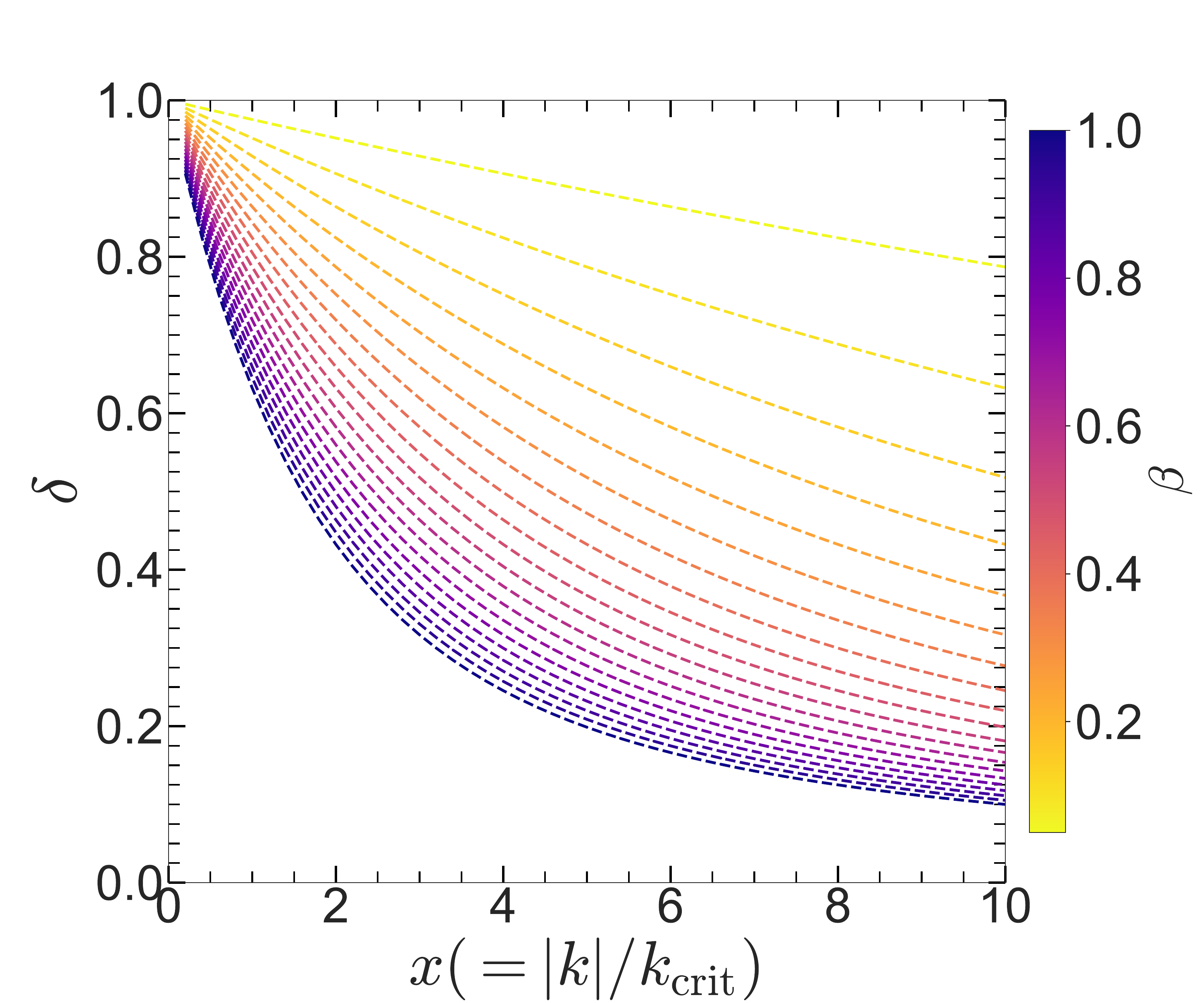}
\caption{Variation of the reduction factor ($\delta$) as a function of dimensionless wavenumber ($x$) are shown for different disc thickness ($\beta$ values). The values of $\beta$ are colour-coded here.}
\label{fig:variation_reduc}
\end{figure}

Fig.~\ref{fig:variation_reduc} shows the variation of the reduction factor, $\delta$ (Eq.~\ref{eq:dimen_delta}) with the dimensionless wavenumber $x$, for various disc thickness (or $\beta$ values). The reduction due to the disc thickness becomes severe for larger values of disc thickness, as expected. Also, the reduction factor is seen to become progressively more important in the short-wavelength (large $x$)
 branch, as expected from the form of Eq.~\ref{eq:dimen_delta}.

\end{appendix}

\end{document}